\documentclass[journal]{IEEEtran}

\usepackage[utf8]{inputenc}
\usepackage{blindtext}
\usepackage{graphicx}
\usepackage{color}
\usepackage{hyperref}

\usepackage[noadjust, nospace]{cite}

\ifCLASSINFOpdf
  \usepackage{graphicx}
\else
\fi
\usepackage[cmex10]{amsmath}
\usepackage{bbm}
\usepackage{amssymb}
\usepackage{algorithmic}

\usepackage{array}
\usepackage{url}

\makeatletter
\@ifclasslater{IEEEtran}{2012/12/26}
 {\newcommand{\killproofname}{\unskip\nopunct}}
 {\newcommand{\killproofname}[1]{\unskip\aftergroup\ignorespaces\ignorespaces}}
\makeatother

\usepackage{graphicx}
\usepackage{subcaption}
\usepackage{subfiles}
\usepackage{amsfonts}
\usepackage{todonotes}
\usepackage{textcomp}
\usepackage{gensymb}
\usepackage{cleveref}

\usepackage{amsthm,amsmath}
\usepackage{bm}
\let\labelindent\relax
\usepackage{enumitem}

\newtheorem{thm}{Theorem}
\newtheorem{lemma}{Lemma}

\theoremstyle{definition}
\newtheorem{definition}{Definition}

\newtheorem{manualtheoreminner}{Lemma}
\newenvironment{manualtheorem}[1]{%
  \renewcommand\themanualtheoreminner{#1}%
  \manualtheoreminner
}{\endmanualtheoreminner}
\makeatletter
\@addtoreset{proofpart}{thm}
\makeatother

\usepackage{algorithm}

\def\dvbar#1{\Vert#1\Vert}

\def\norm#1{\dvbar{#1}}

\begin{document}
%
\title{Sampling and Reconstruction of Bandlimited Signals with Multi-Channel Time Encoding}
%
%
%
\def\AFF{Karen~Adam, Adam~Scholefield and Martin~Vetterli are with the School of Computer and Communication Sciences, Ecole Polytechnique F\'{e}d\'{e}rale de Lausanne (EPFL), CH-1015 Lausanne, Switzerland, email: firstname.lastname@epfl.ch.}

\def\FUND{This work was in part supported by the Swiss National Science Foundation grant number 200021\_181978/1, ``SESAM - Sensing and Sampling: Theory and Algorithms''.}
\author{
Karen~Adam,~\IEEEmembership{Student Member,~IEEE,}
Adam~Scholefield,~\IEEEmembership{Member,~IEEE,}
Martin~Vetterli,~\IEEEmembership{Fellow,~IEEE}
\thanks{\FUND}\thanks{\AFF}}

\maketitle

\begin{abstract}
    Sampling is classically performed by recording the amplitude of an input signal at given time instants; however, sampling and reconstructing a signal using multiple devices in parallel becomes a more difficult problem to solve when the devices have an unknown shift in their clocks.
    
    Alternatively, one can record the times at which a signal (or its integral) crosses given thresholds. This can model integrate-and-fire neurons, for example, and has been studied by Lazar and T\'oth under the name of ``Time Encoding Machines''. This sampling method is closer to what is found in nature.

    In this paper, we show that, when using time encoding machines, reconstruction from multiple channels has a more intuitive solution, and does not require the knowledge of the shifts between machines. We show that, if single-channel time encoding can sample and perfectly reconstruct a $\mathbf{2\Omega}$-bandlimited signal, then $\mathbf{M}$-channel time encoding with shifted integrators can sample and perfectly reconstruct a signal with $\mathbf{M}$ times the bandwidth.
    Furthermore, we present an algorithm to perform this reconstruction and prove that it converges to the correct unique solution, in the noiseless case, without knowledge of the relative shifts between the integrators of the machines. This is quite unlike classical multi-channel sampling, where unknown shifts between sampling devices pose a problem for perfect reconstruction.
\end{abstract}


%
\IEEEpeerreviewmaketitle

\section{Introduction}

    Almost all current sampling theories represent a signal using (time, amplitude) pairs. However, this is quite different from the way encoding is done in nature, where processes have undergone millions of years of evolution.
    More precisely, when a neuron takes an input, it outputs a series of action potentials---the \textit{timings} of which encode the original input.
    
\begin{figure*}[tb]
	\begin{minipage}[b]{\linewidth}
		\centering
		\centerline{\includegraphics[width=\columnwidth]{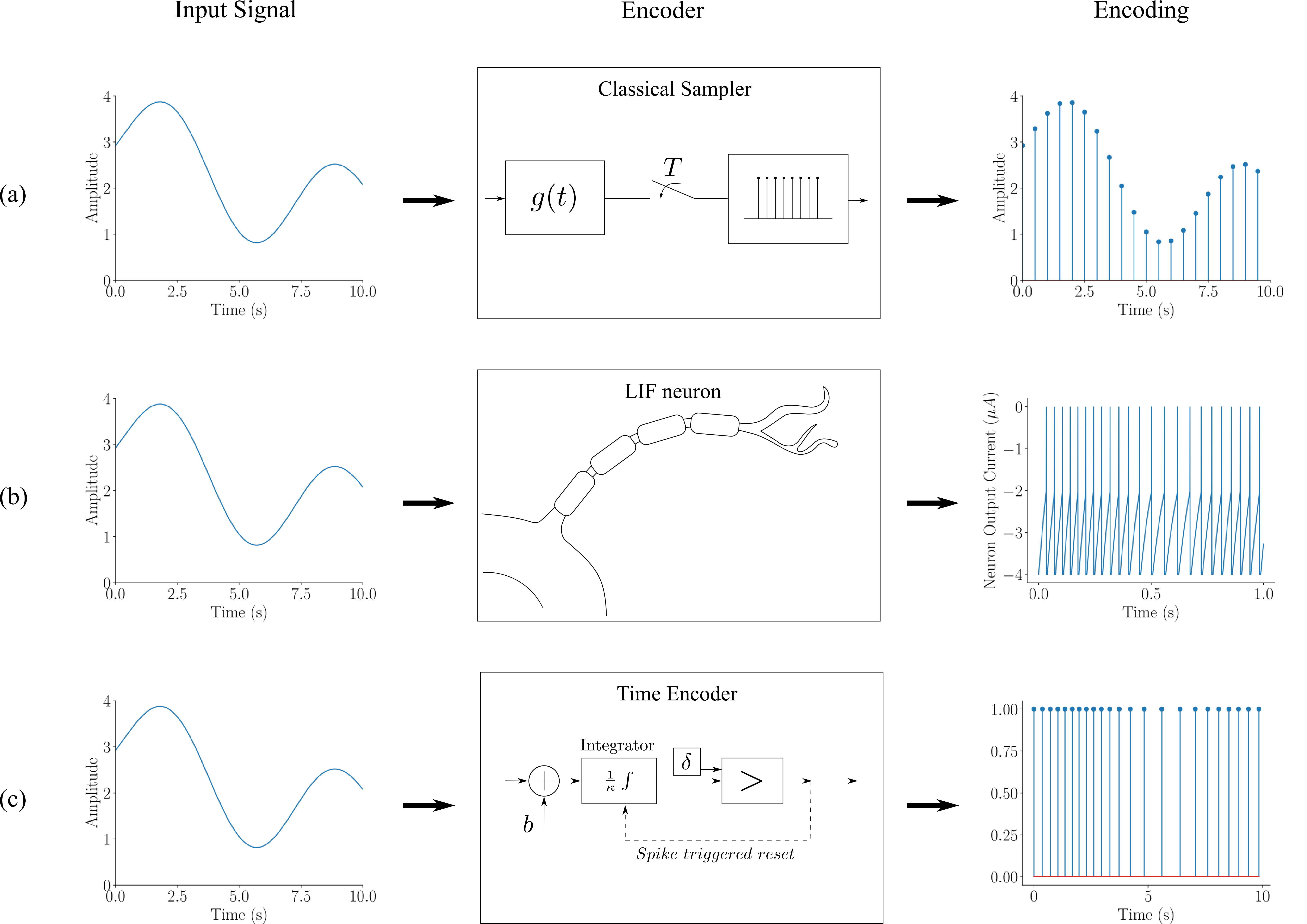}}
	\end{minipage}
	\caption{\label{fig: Figure 1}Encoding of the same signal using different encoding modalities: \textbf{(a)} a classical sampler, \textbf{(b)} a leaky-integrate and fire (LIF) neuron, and \textbf{(c)} a time encoding machine. In \textbf{(a)}, the signal is convolved with a kernel $g(t)$, commonly assumed to be a sinc function if the input signal is bandlimited for example, and then sampled every $T$ seconds. The output is then a series of equally spaced (time, amplitude) pairs. In \textbf{(b)}, we assume that the signal is injected as a current into a spiking leaky-integrate-and-fire neuron following the model described in~\cite{burkitt2006review1} and implemented using a spiking neural network simulator called  Brian~\cite{goodman2009brian}. The recorded output is the outgoing current which exhibits a series of action potentials, or spikes. In \textbf{(c)}, the signal is input to a time encoding machine, as will be described in Section~\ref{sec: Prob Setting}, and the output is a series of signal-dependent trigger times. Notice how, in both \textbf{(b)} and \textbf{(c)}, the output spike streams are denser when the signal is stronger and sparser when the signal is weaker.}
	\end{figure*}

    Similarly, the output of a time encoding machine (TEM)~\cite{lazar2004perfect} is not a series of (time, amplitude) pairs as in classical sampling, but rather a series of signal-dependent time points, which is reminiscent of the output of spiking neurons. The resemblance is highlighted in Fig.~\ref{fig: Figure 1} where we depict the outputs of three different encoding schemes: encoding using classical sampling, using a leaky integrate-and-fire neuron and using a time encoding machine.

    Classical sampling and reconstruction~\cite{shannon1949communication, vetterli2014foundations} has been revisited frequently with extensions proposed along two axes: the sampling setup and the signal class. For example, the traditional uniform sampling setup has been extended to setups where samples are irregularly spaced in time~\cite{feichtinger1994theory, jaffard1991density}, and where samples are taken at unknown locations~\cite{elhami2018sampling, kumar2015bandlimited}, among others. On the other hand, reconstructibility results have been established for multiple signal classes, from bandlimited signals~\cite{shannon1949communication}, to signals in general shift-invariant subspaces~\cite{unser2000sampling}, and signals of finite rate of innovation~\cite{vetterli2002sampling,pan2017towards}. In short, classical sampling is well established and understood. 
    Nonetheless, transitioning to a time encoding paradigm presents additional advantages.
    
    On the one hand, time encoding can help us to better understand biology. Time encoding machines can be made to resemble biological neurons to different degrees. Here, we study perfect integrators that reset once a threshold is reached, but one can also investigate encoding and decoding using leaky integrate-and-fire neurons with refractory periods~\cite{lazar2005multichannel} or even Hodgkin Huxley neurons~\cite{lazar2010population} for more biological resemblence. One can then hope that understanding time encoding can help to better understand the neural code. Moreover, and perhaps more practically, neural networks are often constructed using spiking neurons~\cite{maass1997networks}. Then, understanding the basic components in spiking neural networks can help us understand their functioning and their constraints, as well as understand how to better take advantage of neuromorphic hardware~\cite{davies2018loihi, liu2019event}.
    
    On the other hand, time encoding can help us to improve man-made systems. In fact, time encoding can help us in designing higher-precision sampling hardware as high-precision clocks are more readily available than high-precision quantizers~\cite{lazar2004perfect}. It can also help in reducing power consumption. It has been shown that single-channel time encoding has similar capabilities as traditional sampling:  with time encoding, one can sample and reconstruct bandlimited signals~\cite{lazar2004perfect,rastogi2011integrate,feichtinger2012approximate,saxena2014analyzing} as well as signals with finite rate of innovation~\cite{alexandru2019reconstructing}. Here, we will show that time encoding also provides an advantage over classical sampling when it comes to multi-channel encoding. Indeed, we show that, in time encoding, reconstruction from multi-channel sampling with unknown initial conditions is no harder than reconstruction using single-channel sampling. This is not the case in classical sampling.

	In this paper, we study multi-channel time encoding, where a bandlimited signal is input to $M>1$ time encoding machines that generate different outputs because of a shift in their integrator values. To make the analogy with neuroscience, it seems intuitive (at least from advances in machine learning), that multiple neurons can encode a signal better than one. Here, we would like to quantify this improvement when using TEMs which resemble neurons with perfect integrators. 
	
	Multi-channel sampling has been studied in the classical sampling setup by Papoulis who showed that a bandlimited signal can be reconstructed from its samples from $M$ channels using $1/M$ the sampling rate~\cite{papoulis1977generalized}.

    In this paper, we show that, if a bandlimited signal with bandwidth $\Omega$ can be reconstructed using one TEM, then, using a Projection onto Convex Sets (POCS) algorithm~\cite{sezan1982image, yeh1990iterative}, a bandlimited signal with bandwith $M\Omega$ can be reconstructed from $M$ TEMs with the same parameters, as long as the machines are shifted with nonzero shifts. We also show that the reconstruction algorithm and conditions do not require the knowledge of the shifts, as long as these are nonzero. This is an important improvement over~\cite{adam2019multi-channel}, where we only showed that this bound could be achieved if shifts between the machines were equally spaced, which is not easy to achieve in practice. The bound we propose here generalizes to all shift configurations.

\section{Background and Motivation}

\subsection{Previous Work on Multi-Channel Time Encoding}

As we previously mentioned, time encoding can mimick sensory information processing in neuroscience. Therefore, an intuitive extension to the time encoding machine introduced in~\cite{lazar2004perfect} is a system consisting of multiple time encoding machines: human sensory systems are comprised of many neurons that encode inputs using spikes, which are later used in higher order processes in the brain. 

Moreover, neurons in sensory systems often develop receptive fields.  This means that certain neurons are sensitive to certain shapes of inputs, and the spiking output of these neurons is essentially driven by filtered versions of the original input signal~\cite{hubel1962receptive}. Therefore, different neurons spike at different times and therefore encode different sets of information.  

Inspired by such experimental findings, Lazar and Pnevmatikakis defined a setup with $M$ linearly independent filters and $M$ leaky integrate-and-fire time encoding machines~\cite{lazar2008faithful}. A 1-dimensional signal $x(t)$ is then fed into filter $i$ before being input to TEM $i$ for $i=1\cdots M$ and then reconstructed from the samples. Within this setup, the authors were able to quantify the improvement one obtains from the multi-channel encoding and decoding setup.

In the present paper, our approach to time encoding is different: we assume that we are dealing with multiple \emph{similar} neurons encoding the same input, i.e. they all respond to the same kinds of stimuli.

Therefore, we assume that our signal is not prefiltered before being input to each machine or neuron, but that machines output different spike times because of different initial configurations of the time encoding machines or neurons. In~\cite{adam2019encoding}, we extend the results of this paper to time encoding and decoding of vectors of inputs where the connection between the inputs is more complex. This extension mimicks the way  neurons automatically form receptive fields: these fields arise naturally because of the structure of the connection between input and  neuron.

The setup presented here also draws a parallel with the multi-channel sampling setup in the classical sampling scenario, where sampling devices have unknown shifts in their clocks. Here, our time encoding machines will have unknown shifts in their integrators. The former problem seems to be quite difficult to solve, whereas the latter seems no harder to solve than the single-channel variant.

\subsection{Foundations for understanding time encoding}
The theory behind time encoding and decoding is built on the seminal work on reconstruction from averages presented in~\cite{aldroubi2002non, sun2002reconstruction}. The authors set the foundations for the reconstruction of bandlimited signals from averages. They provided an iterative algorithm for the reconstruction of input signals, with guarantees for convergence under constraints on the number of samples taken. The jump to time encoding and decoding included two extra contributions, the first is the relationship between spike times and average values, and the second is the development of a closed form algorithm that bypasses some impracticalities of the recursive algorithm where the results depend on the number of iterations or on the stopping criterion of the algorithm.

The work in the present paper is heavily inspired by this previous work on reconstruction from averages and on time encoding and decoding. However, we use a POCS approach, which provides geometric intuition, and allows us to  construct a convergence proof which we believe is more approachable. During the revision of this manuscript, a preprint of parallel work about POCS and time encoding became available online~\cite{thao2019pseudoinversion}.\\

\subsection{Multi-channel Time Encoding: Advantages}
Time encoding and classical sampling require similar sampling rates to be able to reconstruct a bandlimited signal. However, as we briefly mentioned, time encoding hints towards hardware that is more power-efficient when compared to traditional analog-to-digital converters (ADC)~\cite{chen2006asynchronous, tang2013continuous}. In fact, time encoding requires only the distinction of two levels to be able to detect the occurrence of a spike, whereas classical ADC requires higher precision quantization. Moreover, time encoding, when based on an integrate-and-fire paradigm, as it is here, produces sparser activity when the input signal has a lower intensity. Therefore, if dealing with signals that are zero for long stretches of time, a time encoder that has no bias will not spike when the signal is zero, thus saving energy\footnote{Note that we assume, in the setup of this paper, that time encoders do have a positive bias. Therefore, these time encoders always spikes, even if the signal is zero. This allows us to ensure theoretical guarantees for reconstruction. More general setups which can admit a bias of zero are, for example, presented in~\cite{adam2019encoding}.}.

Moreover, multi-channel time encoding provides a more scalable alternative to single-channel time encoding with high spiking rate. It is true that a single-channel TEM and a multi-channel TEM have, in theory, the same performance if they have the same total spiking rate, as we show later. However, multi-channel time encoding provides practical advantages:

\begin{enumerate}
    \item When designing hardware for multi-channel time encoding, one can simply reuse single-channel TEMs that have already been well engineered to have a specific spiking rate. On the other hand, modifying the spiking rate of a TEM without perturbing other parameters is not easy. In fact, when designing hardware for time encoding, one often has multiple conflicting specifications to achieve: the integrator circuit should behave as closely as possible to a perfect integrator, the frequency response of the system should be high enough compared to the input signal, the time encoder should be able to achieve a given spiking rate, the spikes should be narrow enough etc. As these characteristics are often conflicting, it is easier to compromise on one of them, namely the spiking rate in this case, in order to satisfy other constraints, and then stack different machines together to obtain a better performing system.
    \item Multi-channel time encoding allows a lower spiking rate per channel. This  allows spikes to be resolvable when they have a nonzero width, which is always the case in practice. Thus, having multi-channel time encoding allows for a better resolution of the spikes for each channel and an overall better estimate of the output spike times.
\end{enumerate}

   \section{Single-Channel TEM}
   \label{sec: Prob Setting}
   \subsection{Time Encoding Definition}

\begin{figure}[tb]
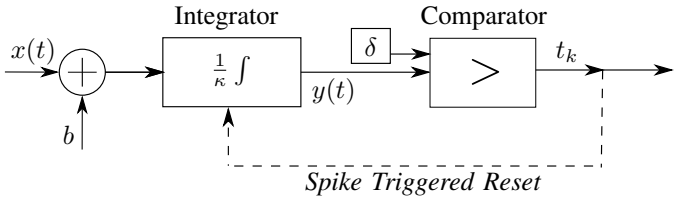

	\begin{minipage}[b]{1.0\linewidth}
		\centering
        \def\svgwidth{\columnwidth}
        \subfile{figures/TEMCircuit.tex}
	\end{minipage}
	\caption{Circuit of a time encoding machine with input $x(t)$ and parameters $\kappa$, $\delta$ and $b$, where $\kappa$ is the integrator constant, $\delta$ is a threshold above which a spike is triggered and $b$ is a positive bias added to the signal.}
	\label{fig: TEM Circuit}
	\end{figure}

\begin{figure}[tb]
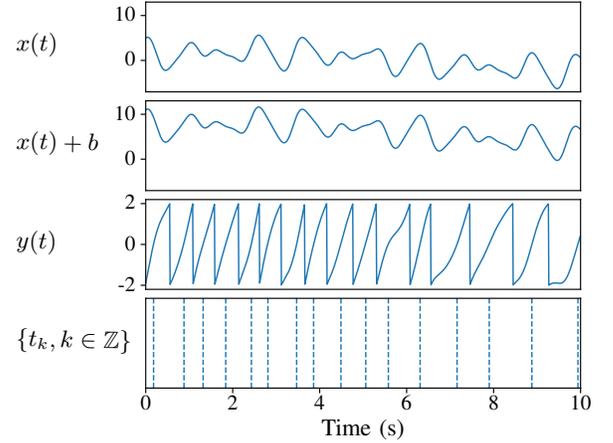

	\begin{minipage}[b]{1.0\linewidth}
		\centering
        \def\svgwidth{\columnwidth}
        \subfile{figures/Figure3_TEExample.tex}
	\end{minipage}
	\caption{Processing of a signal $x(t)$  as it goes through the different stages of a time encoding machine. From top to bottom, we have: the input signal $x(t)$; the result of the bias addition where $b$ is the bias; the result of the integration; and the spike stream output where $t_k$ denotes the $k^{\textrm{th}}$ spike and $k \in \mathbb{Z}$. Note that, in practice and in our simulations, $k$ takes a finite number of values as there is a finite number of spikes, but the analysis in this paper is conducted for infinite number of spikes, assuming the TEM runs forever.}
	\label{fig: Time Encoding Example}
	\end{figure}

    There are different variations of time encoding~\cite{lazar2004perfect,lazar2004timerefractory,lazar2005multichannel, gontier2014sampling}, but we consider the case where a time encoding machine (TEM) acts like an integrate-and-fire neuron with a perfect integrator and no refractory period\footnote{Note that this is slightly different from the definition provided in~\cite{lazar2004perfect}.}.
    \begin{definition}
    \label{def: time encoding machine}
    A \textit{time encoding machine} (TEM) with parameters $\kappa$, $\delta$, and $b$ takes an input signal $x(t)$, adds to it a bias $b$, and integrates the result, scaled by $1/\kappa$, until a threshold $\delta$ is reached. Once this threshold is reached, a time is recorded, the value of the integrator resets to $-\delta$ and the mechanism restarts. We say that the machine spikes at the integrator reset and call the corresponding recorded time $t_k$ a \textit{spike time}. 
	\end{definition}
	Figure~\ref{fig: TEM Circuit} depicts the circuit of a TEM and Fig.~\ref{fig: Time Encoding Example} provides an example of how an input generates its output.
		
	Note how our definition of a sample has changed. In traditional sampling, a sample denoted a (time, amplitude) pair, whereas here, a sample denotes a spike time. We use the terminology ``spike time'', to keep the analogy with integrate-and-fire neurons which produce responses by emitting action potentials. These action potentials have a fixed shape and amplitude\footnote{There are two types of action potentials: the all-or-none action potential has a fixed amplitude, whereas the graded action potential can have a varying amplitude. The integrate-and-fire model assumes that action potentials are all-or-none, and not graded, although graded action potentials can also be found in biology.}, so the relevant information in a neuron's output lies in the \textit{timing} of these action potentials, or spikes.

    Before we proceed, notice that the integrator constant $\kappa$ and the threshold $\delta$ in Definition~\ref{def: time encoding machine} can be combined into one parameter $\delta'=\kappa\delta$. However, we prefer to keep the two parameters separate as each comes from a different source in hardware. The integrator constant $\kappa$ arises from the integrator circuit as depicted in Fig.~\ref{fig: Integrator circuit} and is therefore hard to change. The threshold, on the other hand, is a parameter of the comparator and is easier to manipulate.
	
	\subsection{Iterative Reconstruction of Bandlimited Signals}
	Results on signal reconstruction from the output of a TEM have been obtained for cases where the input is a $c$-bounded, $2\Omega$-bandlimited  signal in $L^2(\mathbb{R})$.
	\begin{definition}
	A signal $x(t)$ is \textit{$2\Omega$-bandlimited} and \textit{$c$-bounded} if its Fourier transform is zero for $|\omega|>\Omega$ and $|x(t)| \leq c$ where $c\in\mathbb{R}$.
	\end{definition}
	\begin{definition}
	A signal $x(t)$ is in $L^2(\mathbb{R})$ if $\int_{-\infty}^{\infty} |x(t)|^2 dt = d <\infty$ for some $d \in \mathbb{R}$.
	\end{definition}

	Although these requirements are generally not met by real-world signals, we adopt the bandlimited constraint as is done in the classical sampling community, assuming that signals generally have a frequency region where most of the interesting information lies~\cite{slepian1976bandwidth}.

    It was  shown by Lazar and T\'oth that such a signal can be perfectly reconstructed from the samples obtained from a TEM with parameters $\kappa$, $\delta$, and $b$, if $b>c$ and
	\begin{equation}
	    \label{eq: bandwidth_single}
	    \Omega < \frac{\pi\left( b-c \right)}{2\kappa\delta}.
	\end{equation}

    The reconstruction algorithm uses the spike times $t_k$ to compute integrals of the original signal~\cite{lazar2004perfect}.
	Indeed, if $x(t)$ is our input signal and $\left\{ t_k, k\in \mathbb{Z}\right\}$ is the set of spike times recorded by our TEM, then we can compute
	\begin{equation}
	\int_{t_k}^{t_{k+1}} x(u) \, du = 2\kappa\delta -b\left(t_{k+1}-t_k\right),\label{eq:integrals_from_trigger_times}
	\end{equation}
	where $t_k$ and $t_{k+1}$ are any two consecutive trigger times. Now, let $\mathcal{R}$  be the following reconstruction operator:
	\begin{equation}
	\label{eq: Rec_operator}
	\mathcal{R}\left(y(t)\right) = \sum_{k \in \mathbb{Z}} \int_{t_k}^{t_{k+1}} y(u) \, du \ \ g(t-s_k),
	\end{equation}
	where $s_k=\left(t_k+t_{k+1}\right)/2$ and $g(t) = \sin(\Omega t)/(\pi t)$. 
	
	Given this $\mathcal{R}$, one can estimate $x(t)$ iteratively by setting
	\begin{align}
	    x_0 &= \mathcal{R}\left(x\right),\label{eq: Rec base step}\\
	    x_{l+1} &=x_l + \mathcal{R}\left(x-x_l\right). \label{eq: Rec inductive step}
	\end{align}
	
	To prove that the reconstruction algorithm converges if~\eqref{eq: bandwidth_single} is satisfied, one requires a bound on the separation between spike times: we recall that $|x(t)|\leq c$, which, when substituted into~\eqref{eq:integrals_from_trigger_times}, yields
	\begin{align}
	    -c\left(t_{k+1}-t_k\right)&\leq 2\kappa\delta - b\left(t_{k+1}-t_k\right),\nonumber\\
	    t_{k+1}-t_k &\leq \frac{2\kappa\delta}{b-c}.\label{eq:spike_seperation_bound}
	\end{align}
	Then, one can use Bernstein and Wirtinger's inequalities (Lemmas~\ref{lemma: bernstein's inequality} and~\ref{lemma: wirtinger's inequality} of Appendix~\ref{sec: Appendix Former Results}) to prove convergence of the algorithm described in~\eqref{eq: Rec_operator}-\eqref{eq: Rec inductive step}. In short, it is shown in~\cite{lazar2004perfect} that the given algorithm can perfectly reconstruct a $c$-bounded, $2\Omega$-bandlimited signal in $L^2(\mathbb{R})$  from the samples of a TEM with parameters $\kappa$, $\delta$ and $b$, given that $c<b$ and $\Omega$ satisfies~\eqref{eq: bandwidth_single}.

	Notice that this result imposes a Nyquist-like constraint on the bandwidth: The bound in ~\eqref{eq: bandwidth_single} requires a bandwidth which is inversely proportional to the separation between spike times. Reconstruction of the original signal is then very similar to the reconstruction of a bandlimited signal sampled with irregularly spaced amplitude samples~\cite{feichtinger1994theory}.
	
	\subsection{Matrix Formulation of Bandlimited Signal Reconstruction}
	In~\cite{lazar2004perfect}, Lazar et al. also obtain a closed-form matrix formulation for the above recursive algorithm. First, let $\mathcal{G}$ be the operator defined as
	\begin{equation*}
	    \mathcal{G}\left(y\right) = \sum_{k \in \mathbb{Z}}y_k g(t-s_k),
	\end{equation*}
	where $s_k=\left(t_k+t_{k+1}\right)/2$ and $g(t) = \sin(\Omega t)/(\pi t)$ as before.
    In addition, define \[\mathbf{q}=\left[\int_{t_k}^{t_{k+1}}x(u)\, du\right]_{k \in \mathbb{Z}},\] and \[\mathbf{H}= {\left[H_{\ell k}\right]_{\ell,k \in \mathbb{Z}} = \left[\int_{t_\ell}^{t_{\ell+1}}g(u-s_k) \, du \right]_{\ell,k \in \mathbb{Z}}}.\] Then, one can write $x(t) = \mathcal{G} \mathbf{H}^+\mathbf{q}{\left(\mathbf{H}^+\mathbf{q}\right)}$ where  $
	\mathbf{H}^+$ is the pseudoinverse of $\mathbf{H}$. We refer the reader to~\cite{lazar2004perfect} for a proof. 
	
	We have covered the main results established in~\cite{lazar2004perfect} and now wish to reformulate the reconstruction algorithm from the perspective of projections onto convex sets. We will later use this perspective to present a solution for multi-channel sampling and reconstruction.

\section{Single-Channel TEM: A POCS Perspective}

	We wish to reach a more intuitive interpretation of the recursive algorithm presented above, to adapt it to new, potentially more complex scenarios. 
    To do so, we will slightly modify the reconstruction algorithm to be able to adopt a projection onto convex sets approach.
	\begin{definition}
	The \textit{projection onto convex sets (POCS) method} obtains a solution for $x$, called $\hat{x}$, by alternately projecting on each of the convex sets $\mathcal{C}_1, \mathcal{C}_2,\cdots,\mathcal{C}_N$, using operators $\mathcal{P}_1,\mathcal{P}_2,\cdots,\mathcal{P}_N$. Here, we assume that $\exists N \in \mathbb{N}$ such that the element $x$ we are looking for lies in the intersection of $N$ known convex sets $\mathcal{C}_1, \mathcal{C}_2,\cdots,\mathcal{C}_N$ which are subsets of a Hilbert space $X$. 
	\end{definition}
	The POCS algorithm is known to converge to a fixed point which lies in the intersection of the sets at hand $\bigcap_{i=1}^N\mathcal{C}_i $~\cite{bauschke1996projection}. Thus, if the intersection of the sets consists of a single element, then the algorithm converges to the \textit{correct} solution.
	
	One can see that the algorithm presented by Lazar and T\'oth~\cite{lazar2004perfect} resembles a POCS algorithm. In particular, notice that $\mathcal{R}$, defined in~\eqref{eq: Rec_operator} can be rewritten as 
	\begin{equation}
	    \mathcal{R}\left(y(t)\right) = \mathcal{B}\left(y(t)\right) \ast g(t),
	\end{equation}
	where 
	\begin{equation}
	    \mathcal{B}\left(y(t)\right) =\sum_{k \in \mathbb{Z}} \int_{t_k}^{t_{k+1}} y(u) \, du \ \ \delta(t-s_k).
	\end{equation}
	Here, $\delta(t)$ is the Dirac delta. In words, $\mathcal{B}$ adds Diracs in the center of each inter-spike interval\footnote{The inter-spike intervals are the intervals $\left[t_k, t_{k+1}\right]$ between any two consecutive spikes $t_k$ and $t_{k+1}$.}, where the Diracs are weighted in such a way that the input and output have the same integrals between $t_k$ and $t_{k+1}$. 
	 The operator $\mathcal{R}$ first applies $\mathcal{B}$ and then convolves the result with a sinc function $g(t)$ to make it bandlimited. 
	
	Recursively applying $\mathcal{R}$, as in~\eqref{eq: Rec base step} and~\eqref{eq: Rec inductive step}, therefore alternately projects onto two convex sets, the set of bandlimited functions and the set of functions which match the measurements $\left\lbrace t_k, k\in\mathbb{Z}\right\rbrace$ . However, the range of operator $\mathcal{B}$ does not lie in a Hilbert space, so the algorithm does not meet all the technical requirements for a properly converging POCS algorithm~\cite{thao2019pseudoinversion}. 
	To remedy this, 
	we assume that our input signals are in $L^2(\mathbb{R})$, and define operator $\mathcal{B}_1$ as follows:
		\begin{equation}
	    \label{eq: B proxy operator for A}
	    \mathcal{B}_1\left(y(t)\right) =\sum_{k \in \mathbb{Z}} \int_{t_k}^{t_{k+1}} y(u) \, du \ \ \frac{1}{t_{k+1}-t_k}\mathbbm{1}_{[t_k, t_{k+1})}(t),
	\end{equation}
	where $\mathbbm{1}_{[t_k, t_{k+1})}(t)$ is a function which takes value one when $t \in [t_k, t_{k+1})$ and zero elsewhere. 
    $\mathcal{B}_1$ and $\mathcal{B}$ produce signals that have the same integrals over intervals $\left[t_k, t_{k+1}\right)$, but the result obtained from applying $\mathcal{B}_1$ is in $L^2(\mathbb{R})$, which is a Hilbert space.
	
	Now, we define operator $\mathcal{R}_1$ as follows:
	\begin{equation}
	\label{eq: redefining rec op}
	\mathcal{R}_1\left(y(t)\right) =  \mathcal{B}_1\left(y(t)\right) \ast g(t).
	\end{equation}
	
	Now defining
	\begin{equation}
	    x_0 = \mathcal{R}_1\left(x\right), \quad x_{\ell+1} = x_\ell + \mathcal{R}_1(x-x_\ell),
	\end{equation}
	we can show that $x_\ell(t)$ is bandlimited with bandwidth $2\Omega$ at every iteration $\ell$. Therefore,
	\begin{align}
	x_{\ell+1} &= x_\ell\ast g + \mathcal{B}_1\left(x-x_\ell\right) \ast g\notag,\\
	 &= \left(x_\ell + \mathcal{B}_1\left(x-x_\ell\right)\right) \ast g\notag.
	\end{align}.

    Notice the similarity between the iterations of this algorithm and those of the algorithm presented in Section~\ref{sec: Prob Setting}. Earlier, at each iteration, Diracs were placed between consecutive spikes to make the signal consistent with the spike times and the result was then low-pass filtered. Here, indicator functions are placed between consecutive spikes before the low-pass filter is applied.

    To formalise the POCS perspective, we can divide the computation of $x_{\ell+1}$ into two steps:
    \begin{equation}
        x_{\ell+1} = \mathcal{P}_{\Omega}\left( \mathcal{P}_{\mathrm{A}_1} \left(x_\ell\right)\right),
    \end{equation}
    where
    \begin{equation}
    \label{eq: projection onto space of functions with right integral}
    \mathcal{P}_{\mathrm{A}_1}\left(y(t)\right) = y(t) + \mathcal{B}_1(x(t)-y(t)),
    \end{equation}
    and
    \begin{equation}
    \label{eq: projection onto bandlimited function space}
    \mathcal{P}_{\Omega}\left(y(t)\right) = y(t)*g(t).
    \end{equation}

    Letting $\mathcal{C}_{\Omega}$ be the space of $2\Omega$-bandlimited functions which are also in $L^2(\mathbb{R})$, we have the following two lemmas.
    \begin{lemma}
\label{lemma: P2 projection operator}
    $\mathcal{P}_{\Omega}$ is a firmly nonexpansive projection operator onto $\mathcal{C}_{\Omega}$.
\end{lemma}
\begin{proof}
See Appendix~\ref{sec: Appendix Proofs}.
\end{proof}

\begin{lemma}
\label{lemma: C2 convex}
$\mathcal{C}_\Omega$ is convex.
\end{lemma}
\begin{proof}
See Appendix~\ref{sec: Appendix Proofs}.
\end{proof}
    
    As for $\mathcal{P}_{\mathrm{A}_1}$, we can substitute~\eqref{eq: B proxy operator for A} into~\eqref{eq: projection onto space of functions with right integral}, yielding
    \begin{align}
        \mathcal{P}_{\mathrm{A}_1}&\left(y(t)\right) = \notag\\
        &y(t)+ \sum_{k \in \mathbb{Z}} \int_{t_k}^{t_{k+1}} \left[x(u)-y(u)\right] \, du \ \ \frac{\mathbbm{1}_{[t_k, t_{k+1})}(t)}{t_{k+1}-t_k}.
    \end{align}
    We thus see that the operator depends on the spike times $t_k$ emitted by a TEM with input $x(t)$. $\mathcal{P}_{\mathrm{A}_1}$ used operator $\mathcal{B}_1$ to produce an output which is consistent with the measurements $\left\lbrace t_k, k\in\mathbb{Z}\right\rbrace$.

    Now, let $\mathcal{C}_{\mathrm{A}_1}$ be the space of $L^2(\mathbb{R})$ functions $y(t)$ satisfying $\int_{t_k}^{t_{k+1}}y(u)\, du = \int_{t_k}^{t_{k+1}}x(u)\, du, \quad \forall k \in \mathbb{Z}$. In other words, the space $\mathcal{C}_{\mathrm{A}_1}$ is the space of functions in $L^2(\mathbb{R})$ that are consistent with the measurements $t_k$: these functions generate the spike times $t_k$ when passed through the TEM A$_1$.

    \begin{lemma}
\label{lemma: P1 projection operator}
    $\mathcal{P}_{\mathrm{A}_1}$ is a firmly nonexpansive projection operator onto $\mathcal{C}_{\mathrm{A}_1}$.
\end{lemma}
\begin{proof}
See Appendix~\ref{sec: Appendix Proofs}.
\end{proof}

\begin{lemma}
\label{lemma: C1 convex}
$\mathcal{C}_{\mathrm{A}_1}$ is convex.
\end{lemma}
\begin{proof}
See Appendix~\ref{sec: Appendix Proofs}.
\end{proof}

    Since both $\mathcal{P}_{\mathrm{A}_1}$ and $\mathcal{P}_{\Omega}$ are projection operators onto $\mathcal{C}_{\mathrm{A}_1}$ and $\mathcal{C}_{\Omega}$ respectively, the entire iterative reconstruction algorithm then consists of alternately projecting onto two sets, each being convex.

    We have thus devised a  reconstruction algorithm for single-channel time encoding consisting of an alternating projection onto convex sets (POCS) algorithm~\cite{cheney1959proximity,bauschke1996projection,papoulis1975new}.
	
	Adopting a POCS interpretation of our algorithm allows us to directly deduce that the algorithm converges to a fixed point in the intersection of the sets of $2\Omega$-bandlimited functions, and functions which generate the spike times of the TEM. The conditions for the fixed point to be unique and for it to indeed be the original signal relies on our proof given in Section~\ref{sec: Uniqueness}.
	
	Therefore, the reconstruction algorithm presented here and in~\cite{lazar2004perfect} remains heavily based on the reconstruction from averages algorithm provided in~\cite{feichtinger1994theory, sun2002reconstruction, sun2002breconstruction},  but the POCS formulation provides intuition on the process and allows for a different approach to proving convergence.

\section{M-Channel TEM}
\subsection{M-Channel TEM Definition}

First let us define integrator shifts between TEMs with the same parameters.
	\begin{definition}
	$M$ TEMs with parameters $\kappa$, $\delta$ and $b$ have \textit{integrator shifts} $\alpha_1, \alpha_2, \cdots, \alpha_M$ if, for the same input $x(t)$, and for any time $t$, the outputs of the integrators $y_1(t), y_2(t), \cdots,y_M(t)$ satisfy
	\begin{align}
	    y_{i+1}(t) &= (y_i(t)+\alpha_i) \mod 2\delta, \ \ i=1\cdots M-1 \label{eq: integrator relationships 1}\\ 
	    y_1(t) &= (y_M(t)+\alpha_M) \mod 2\delta. \label{eq: integrator relationships 2}
	\end{align}
	Here, the $\alpha_i$'s naturally satisfy \footnote{This arises from recursively expanding the expression of $y_1(t)$ in~\eqref{eq: integrator relationships 2} and noting that $y_1(t) = y_1(t) + \sum_i \alpha_i \mod 2\delta$.} $\left(\sum_i \alpha_i \right)\mod 2\delta = 0$.
	\end{definition}

Now, we can define an $M$-channel time encoding machine.

	\begin{definition} \label{def: MTEM}
	An \textit{$M$-channel time encoding machine} consists of $M$ single-channel TEMs $\mathrm{A}_1, \mathrm{A}_2, \cdots, \mathrm{A}_M$, with parameters $\kappa$, $\delta$ and $b$ and integrator shifts $\alpha_1,\cdots, \alpha_M$.
	\end{definition}
	When these shifts are all nonzero, the machines will spike in this order $\mathrm{A}_1, \mathrm{A}_2, \cdots, \mathrm{A}_M$, i.e.
	\begin{align}
	    t_{k-1}^{(i+1)}&< t_k^{(i)}&<t_k^{(i+1)}& \qquad \forall k, \forall i=1\cdots M-1, \label{eq: spike time order 1}\\
	    t_{k}^{(1)}&<t_k^{(M)}&<t_{k+1}^{(1)} \label{eq: spike time order 2}\ & \qquad \forall k,
	\end{align}
	where $\left\lbrace t_k^{(i)}, k \in \mathbb{Z} \right\rbrace$ is the set of spike times emitted by TEM A$_i$.
	
	Equations~\eqref{eq: spike time order 1} and~\eqref{eq: spike time order 2} force a strict order of the spike times on the $M$ machines, which naturally arises from nonzero shifts between the integrators of the machines.

			\begin{figure}[tb]
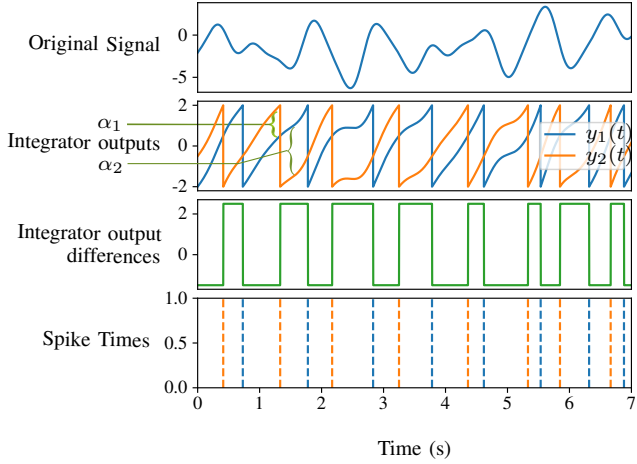

	\begin{minipage}[b]{1.0\linewidth}
		\centering
		\centerline{		        
		\def\svgwidth{\columnwidth}
		\subfile{figures/Figure4_integ_comparison.tex}}
	\end{minipage}
	\caption{Output of the integrators of two TEMs with nonzero shifts. We assume both TEMs have a threshold $\delta = 2$ and that TEM A$_2$ is leading TEM A$_1$ by $\alpha_1=0.75$. This means $y_2(t)= y_1(t)+\alpha_1 \mod 2\delta, \ \forall t$. We plot, from top to bottom: The original signal input to the machines, the output of the integrator of each machine ($y_i(t)$ corresponding to the output of the integrator of TEM A$_i$), the difference between the outputs of the two machines, and the output spikes of each machine. Note how the spike times are interleaved, i.e. there is always one spike on TEM A$_1$ between any two spikes of TEM A$_2$ and vice versa.}
	\label{fig: Integrator Shift example}
	\end{figure}

    Figure~\ref{fig: Integrator Shift example} shows an example of 2-channel time encoding. We pass an input signal through the two single-channel TEMs (with nonzero integrator shifts) and record the output of each integrator. Notice how the integrator values are always shifted by the same amount (modulo $2\delta$). In contrast, if the integrator shifts between the two channels were zero, the output of both integrators would match at all time points. However, in the example presented, the integrators are shifted by a nonzero amount. Therefore, as the spike times are generated at the integrator reset, the TEMs are guaranteed to spike at different times so that $t_k^{(i)}\neq t_\ell^{(j)} \quad \forall k, l \in \mathbb{N}, \quad \forall i\neq j, i, j\in [1,\cdots M]$. Moreover, the spike times are interleaved, satisfying~\eqref{eq: spike time order 1} and~\eqref{eq: spike time order 2}. Note that these equations are naturally satisfied when we have an $M$-channel TEM where all channels have the same parameters $\kappa$, $\delta$, and $b$ and nonzero shifts. However, we have yet to explain where these shifts come from, in practice.

So far, we have assumed that our signals have infinite support, and that our TEM samples the signal for infinite time. In practice, however, a TEM would start recording a signal at a certain time $t_{start}$ and stop recording at $t_{end}$. In these scenarios, integrator shifts can be well defined and implemented.

Indeed, these integrator shifts will result from different initial conditions on the integrators of the TEMs at $t_{start}$. For example, assume TEMs A$_1$ and A$_2$ start integrating at the same time $t_{start}$ with initial values $y_1(t_{start})$ and $y_2(t_{start})$, respectively. Then TEM A$_2$ will always lead TEM A$_1$ by $\alpha_1 = y_2(t_{start})-y_1(t_{start})$.

\begin{figure}[tb]
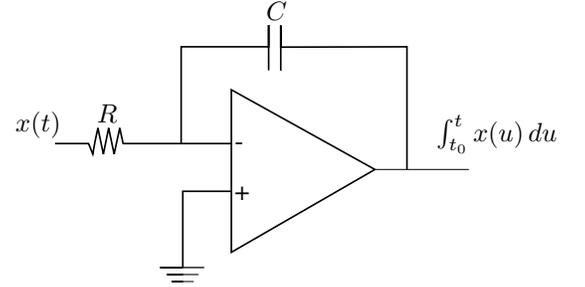

	\begin{minipage}[b]{1\linewidth}
		\centering
		\centerline{		        
		\def\svgwidth{0.74\columnwidth}
		\subfile{figures/opamp.tex}}
	\end{minipage}
	\caption{The circuit of an integrator comprises of an operational amplifier, a resistor $R$ and a capacitor $C$ in the shown configuration. The circuit does not provide a perfect integrator as we require in our model but it serves a good approximation of it and allows the implementation of our setup in hardware. An analysis of time encoding using leaky integrators such as this one is presented in~\cite{lazar2008faithful}.}
	\label{fig: Integrator circuit}
	\end{figure}

More practically, an integrator is represented in circuitry by an operational amplifier coupled with a resistor and capacitor, as seen in Fig.~\ref{fig: Integrator circuit}. This capacitor can be charged with a certain voltage before the input is fed into the circuit. This initial charge of the capacitor can practically implement the initial value of the integrator. Therefore, having different initial charges on the capacitors of each machine would lead to nonzero integrator shifts.

However, we recall that our setup assumes a perfect integrator and infinite time support. The initial conditions formulation in this section serves as a more intuitive explanation of how integrator shifts arise, and as a practical explanation of where these shifts come from, in hardware.

\subsection{Convergence of M-Channel Reconstruction using POCS}
\label{sec: POCS}

We use the POCS formulation to devise a reconstruction algorithm  for multi-channel time encoding, thus extending the result found in~\cite{lazar2004perfect} to the case where $x(t)$ is sampled using multiple independent time encoding machines.

	In~\cite{adam2019multi-channel}, we presented an algorithm that could reconstruct $2\Omega$-bandlimited signals from an $M$-channel TEM if
	\begin{equation}
	    \Omega< \frac{\pi(b-c)}{2\kappa\bar{\alpha}}.
	\end{equation}
	Here, $\bar{\alpha} = \max_{i=1\cdots M}\alpha_i$ depends on the shifts between the machines. In this scenario, we reached the maximal possible bandwidth if the machine integrators were spaced in such a way that $\bar{\alpha}=\alpha_i=2\delta/M, \ \forall i=1\cdots M$. Then, the bandwidth could improve by a factor of $M$ compared to the single channel case.
	
	In this paper, we want to show that, in the $M$-channel case, the improvement on $\Omega$ is always $M$-fold, regardless of the spacing between the machines' integrators, as long as this spacing is nonzero.
    
    To do this, we will design an algorithm that reconstructs an input from its $M$-channel spiking output and provide conditions for its convergence. We use as inspiration the POCS intepretation of the single-channel reconstruction algorithm.

	The POCS method can guarantee convergence onto a fixed point by alternately projecting onto convex sets. The averaged projection method works similarly.
	\begin{definition}
	The \textit{averaged projections method} assumes that we have $N$ convex sets $\mathcal{C}_1,\cdots,\mathcal{C}_N$ with corresponding projection operators $\mathcal{P}_1,\cdots,\mathcal{P}_N$ and that we compute an estimate of $x$ at iteration $\ell+1$ by taking
	\begin{equation}
	    x_{\ell+1} = \frac{1}{N}\sum_{i=1}^{N} \mathcal{P}_i\left(x_\ell\right).
	\end{equation}
	\end{definition}
	This algorithm can be reduced into an alternating projection algorithm and therefore also converges to a fixed point in the intersection of the sets.

	We use the averaged projections method to design an algorithm for reconstructing $2\Omega$-bandlimited signals from the encoding of more than one TEM. Our  algorithm will converge to a fixed point in the set of solutions that can produce the different time encodings, as is guaranteed by the properties of the POCS algorithm and the averaged projections method. Indeed, sufficient technical requirements for the convergence of our algorithm to a fixed point in $\bigcap_{i=1}^N\mathcal{C}_i $ are based on a review by Bauschke and Borwein~\cite{bauschke1996projection}.
	
	Later, we will find conditions on $\Omega$ that are sufficient for this set of solutions to consist of a single element, so that our algorithm converges to the unique and desired solution. First, let us explain the algorithm.
	
	\begin{figure}[tb]
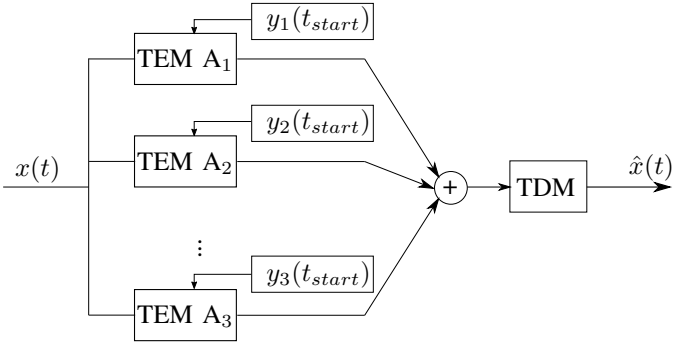

	\begin{minipage}[b]{1.0\linewidth}
		\centering
		\centerline{
		        \def\svgwidth{\columnwidth}
		    \subfile{figures/mtem.tex}}
	\end{minipage}
	\caption{Multi-channel time encoding and decoding pipeline. In Practice, the TEMs are initialized with some initial value of their integrators $y_i(t_{start})$ and the integrator shift between two machines A$_i$ and A$_{i+1}$ is $\alpha_i = y_{i+1}(t_{start})-y_i(t_{start}) \mod 2\delta$, for $1<i<M$ (the shift between machines A$_M$ and A$_1$ is $\alpha_M = y_{1}(t_{start})-y_M(t_{start}) \mod 2\delta$). The output streams of the different machines can be combined into one before being fed into a single decoding machine because of the perfect ordering of the spikes provided in~\eqref{eq: spike time order 1} and~\eqref{eq: spike time order 2}.}
	\label{fig: M-Channel TEM}
	\end{figure}

	Let $\mathrm{A}_1, \mathrm{A}_2, \cdots, \mathrm{A}_M$ be our $M$ time encoding machines, and let $\left\{ t_k^{\left(i\right)}, k \in \mathbb{Z}\right\}$ be the spike times emitted by machine $i$, $i=1 \cdots M$, when the input is $x(t)$---a $2\Omega$-bandlimited signal in $L^2(\mathbb{R})$ such that $|x(t)|<c$, for some $c\in\mathbb{R}$.
	
	Then, let $\mathcal{R}_i$ be the reconstruction operator associated with machine $i$, such that
	\begin{equation}
		\label{eq: Rec operator single channel}
	    \mathcal{R}_i \left(x(t)\right) = \sum_{k \in \mathbb{Z}} \int_{t_k^{(i)}}^{t_{k+1}^{(i)}} x(u) \, du \ \ \frac{\mathbbm{1}_{[t_k, t_{k+1})}(t)}{t_{k+1}-t_k} \ast g\left(t\right). \\
	\end{equation}
    Note that each of these reconstruction operators $\mathcal{R}_i$ is the same as the operator defined in~\eqref{eq: redefining rec op} for each of the individual machines A$_i$, and therefore also consists of two projections onto convex sets $\mathcal{C}_{A_i}$ and $\mathcal{C}_\Omega$.
	
	Then, define a new reconstruction operator
	\begin{equation}
	\label{eq: Rec operator M channel}
	    \mathcal{R}_{1\cdots M} = \frac{1}{M} \sum_{i=1}^{M} \mathcal{R}_i
	\end{equation}
	and recursively estimate $x(t)$ by setting
	\begin{align}
	    \label{eq: Rec Alg M channel base step}
	    x_0 &= \mathcal{R}_{1\cdots M}\left(x\right),\\
	    \label{eq: Rec Alg M channel inductive step}
	    x_{\ell+1} &= x_\ell + \mathcal{R}_{1\cdots M}\left(x-x_\ell\right).
	\end{align}
	
	This algorithm is equivalent to taking alternating projections on the set $\mathcal{C}_{\Omega}$ and the sets $\mathcal{C}_{\text{A}_i}$, where $\mathcal{C}_{\Omega}$ denotes the set of $2\Omega$-bandlimited functions and each $\mathcal{C}_{\text{A}_i}$ denotes the set of functions that could have potentially generated the spike times of machine $\mathrm{A}_i$. All of these sets are convex by Lemmas~\ref{lemma: C2 convex} and~\ref{lemma: C1 convex}.
	
	The algorithm converges to a solution that is $2\Omega$-bandlimited and that can generate the spike times $\left\{t_k^{(i)},k\in\mathbb{Z}\right\}, \ \forall i=1\cdots M$, by the properties of the POCS method. Note that this algorithm does not require knowing the shifts $\alpha_i$ between the integrators of the machines, it only requires knowing the parameters $\kappa$, $\delta$ and $b$ of the machines.
	
	So far, we have only shown that the algorithm converges to a fixed point that satisfies $2\Omega$-bandlimitedness and is consistent with the spike times generated by all machines A$_i$, $i=1\cdots M$.

	\begin{figure}[tb]
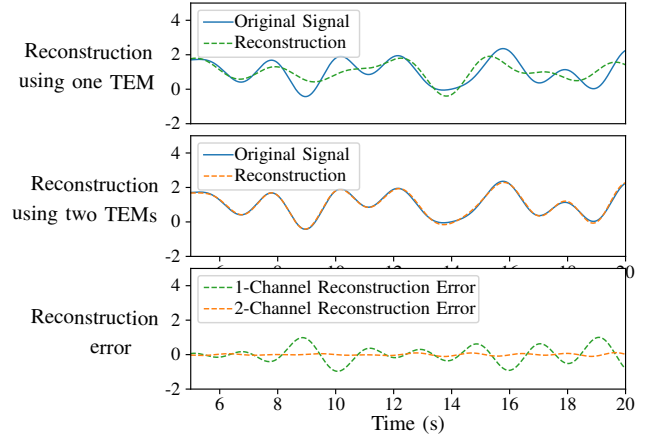

	\begin{minipage}[b]{1.0\linewidth}
		\centering

		\def\svgwidth{\columnwidth}
		\subfile{figures/Figure7_MChannelRec.tex}
	\end{minipage}
	\caption{(Top) Reconstruction of a signal from its time encoding, using one channel. (Middle) Reconstruction of the same signal from its time encoding using two channels with integrators shifted by an unknown value. (Bottom) Reconstruction error when using outputs of 1-channel TEM and 2-channel TEM.}
	\label{fig: M-Channel Rec}
	\end{figure}
	
	In the next section, we show that this solution is unique (and is thus the originally sampled signal), if the signal is $2\Omega$-bandlimited where 
	\begin{equation}
    \label{eq: Omega bound multi}
        \Omega < \frac{M\pi(b-c)}{2\kappa\delta}.
    \end{equation}
    
    We recall that $M$ is the number of machines, $\kappa$, $\delta$ and $b$ are the parameters of the individual machines and $c$ is the bound on the input signal $x(t)$, i.e $|x(t)|\leq c$. 
    Before we proceed to prove uniqueness, we show, in Fig.~\ref{fig: M-Channel Rec}, a reconstruction example demonstrating that the algorithm we suggested for the $M$-channel case can reconstruct a wider range of signals than is possible in the single channel case.

\subsection{Uniqueness of M-Channel Reconstruction using POCS}
\label{sec: Uniqueness}
	We have presented an algorithm which converges to a fixed point in the intersection of the $\mathcal{C}_{\text{A}_i}$'s with $\mathcal{C}_{\Omega}$; we now wish to pinpoint sufficient conditions for this intersection to be unique.
	
	Assume the input signal, $x(t)$, is a $c$-bounded, $2\Omega$-bandlimited signal in $L^2(\mathbb{R})$,  which has its integral well defined: $\forall t,\, \exists \gamma \in\mathbb{R},\, s.t. \int_{-\infty}^{t} x(u)\, du = \gamma$. We wish to find an estimate of the input signal $x(t)$, which we denote $\hat{x}(t)$, using the output spike times of $M$ time encoding machines. By applying the algorithm we described in ~\eqref{eq: Rec operator M channel}-\eqref{eq: Rec Alg M channel inductive step}, $\hat{x}(t) = \lim_{\ell\rightarrow \infty}x_\ell(t)$ will be a fixed point in the intersection of the sets $\mathcal{C}_{\text{A}_i}$ with $\mathcal{C}_{\Omega}$, so $\hat{x}(t)$ will lie in every one of the $\mathcal{C}_{\text{A}_i}$'s. This means that for every $i=1\cdots M$,
	\begin{equation}
	    \int_{t^{(i)}_k}^{t^{(i)}_{k+1}}\hat{x}(u) \, du = \int_{t^{(i)}_k}^{t^{(i)}_{k+1}}x(u)\, du, \quad \forall k\in\mathbb{Z}.
	\end{equation}
	
	Let us denote $X(t) = \int_{-\infty}^t x(u)\, du$ and $\hat{X}(t) = \int_{-\infty}^t \hat{x}(u)\, du$. Then it follows that $X(t^{(i)}_{k+1})-X(t^{(i)}_k)=\hat{X}(t^{(i)}_{k+1})-\hat{X}(t^{(i)}_k), \forall k \in \mathbb{Z}$.
	\begin{lemma}
	$X(t^{(i)}_k)=\hat{X}(t^{(i)}_k),\quad  \forall \ k \in \mathbb{Z}, \quad \forall i=1\cdots M.$
	\end{lemma}
	\begin{proof}
	\begin{align}
        X(t^{(i)}_k) &\stackrel{(a)}{=} \int_{-\infty}^{t^{(i)}_k}x(u) \,du \notag\\
        &\stackrel{(b)}{=} \lim_{k\rightarrow\infty} \sum_{\ell=-\infty}^{k-1} \left(X(t^{(i)}_{\ell+1})-X(t^{(i)}_\ell)\right) \notag \\
        &\stackrel{(c)}{=} \lim_{k\rightarrow\infty} \sum_{\ell=-\infty}^{k-1} \left(\hat{X}(t^{(i)}_{\ell+1})-\hat{X}(t^{(i)}_\ell)\right) \notag \\
        &\stackrel{(d)}{=} \int_{-\infty}^{t^{(i)}_k}\hat{x}(u) \,du \notag \\
        &\stackrel{(e)}{=} \hat{X}(t^{(i)}_k), \quad \forall k \in \mathbb{Z}, \notag
    \end{align}
    where equalities $(a)$ and $(e)$ follow from the definitions of $X(t)$ and $\hat{X}(t)$, respectively,  $(b)$ and $(d)$ follow from $x(t)$ and $\hat{x}(t)$ having well-defined integrals, and $(c)$ follows from the fact that $X(t^{(i)}_{k+1})-X(t^{(i)}_k)=\hat{X}(t^{(i)}_{k+1})-\hat{X}(t^{(i)}_k), \forall k \in \mathbb{Z}$.
    So $X(t)$ and $\hat{X}(t)$ match at all $t^{(i)}_k,\  k\in \mathbb{Z},\  i=1\cdots M$.
	\end{proof}
	
	\begin{lemma}
	The integrals $X(t)$ and $\hat{X}(t)$ are both $2\Omega$-bandlimited.
	\end{lemma}
	\begin{proof}
	The original signals $x(t)$ and $\hat{x}(t)$ are both $2\Omega$-bandlimited. Taking the integrals of these signals corresponds to a division by $j\omega$ in the frequency domain, where $\omega$ denotes the frequency, so the frequency content of $X(t)$ and $\hat{X}(t)$ remains concentrated in $\left[ -\Omega, \Omega\right]$.
	\end{proof}

    Therefore, $X(t)$ and $\hat{X}(t)$ are two $2\Omega$-bandlimited functions which coincide at time points $t_k^{(i)}, \ \forall k\in \mathbb{Z}, \ \forall i=1\cdots M$. In other words, if both $X(t)$ and $\hat{X}(t)$ are sampled at the $t_k^{(i)}$'s, their samples would have the same values.
    
    Let us combine and order all spike times from the machines into one set $\left\{\tilde{t}_k, k \in \mathbb{Z}\right\}$. 
    To show that these samples are sufficient to ensure that $X(t)$ and $\hat{X}(t)$ match, we use a result from Jaffard. In~\cite{jaffard1991density}, he proved that a sampling sequence $\left\{t_k, k\in\mathbb{Z}\right\}$ generates a frame for the space of $2\Omega$-bandlimited functions if and only if $\left\{t_k, k\in\mathbb{Z}\right\}$ is relatively separated and
    \begin{equation}
        \liminf_{r\rightarrow \infty} \frac{n(r)}{r}>\frac{\Omega}{\pi},
    \end{equation}
    where $n(r)$ is the number of samples in an interval of length $r$.

    This finding provides sufficient conditions for irregular (time, amplitude) samples to completely characterize a bandlimited signal: the sample set has to be relatively separated, and the average sampling rate needs to be higher than the Nyquist rate. Note that the set being relatively separated is only required for the reformulation in terms of a problem about frames.
    
    It is a technical condition which is naturally satisfied in our scenario. In fact, it ensures a minimum separation between sample times. In Appendix~\ref{sec: Appendix Proofs}, we formally define it (Definition~\ref{def: relatively separated}) and show that the sampling set $\left\{\tilde{t}_k, k \in \mathbb{Z}\right\}$ is relatively separated (Lemma~\ref{lemma: combined spike times are relatively separated}). 
    On the other hand, to help us prove that the Nyquist-like condition is satisfied, the following lemma provides us with a lower bound on the average sampling rate of our spike times $\left\{\tilde{t}_k, k \in \mathbb{Z}\right\}$.

    \begin{lemma}
    \label{lemma: min multi-channel sampling rate}
    The sampling set $\left\{\tilde{t}_k, k \in \mathbb{Z}\right\}$ has an average sampling rate which is at least $M(b-c)/(2\kappa\delta)$.
    \end{lemma}
    \begin{proof}
    Spike times have a maximal separation between them defined by~\eqref{eq:spike_seperation_bound}. According to this bound, every machine produces a sampling set $\left\{t_k^{(i)}, k \in \mathbb{Z}\right\}$ where two spike times have a separation of at most $2\kappa\delta/(b-c)$. Therefore, the sampling rate $n(r)/r$ is  at least $(b-c)/{2\kappa\delta}$, for any $r\in\mathbb{R}$ . Therefore, the average sampling rate of a machine $\liminf_{r\rightarrow \infty}{n(r)}/{r}$ is at least $(b-c)/2\kappa\delta$. Since all machines fire at distinct time points (because the shifts between them are nonzero), together, they have an average sampling rate which is at least $ {M(b-c)}/{2\kappa\delta}$. 
    \end{proof}

    It follows that the samples emitted by the TEMs are sufficient to determine uniqueness for a $2\Omega$-bandlimited signal, provided that $\Omega$ satisfies~\eqref{eq: Omega bound multi}.

    Hence, a signal $X(t)$ which is $2\Omega$-bandlimited, with $\Omega$ satisfying~\eqref{eq: Omega bound multi}, is uniquely defined by the samples provided by a $M$-channel TEM with parameters $\kappa$, $\delta$ and $b$ and input $x(t)$, such that $|x(t)|\leq c<b$, $x(t) \in L^2(\mathbb{R})$ and has a well defined integral, if the shifts between the machines are nonzero. Therefore, $X(t)$ and its estimate $\hat{X}(t)$ match exactly, and as $x(t)$ and $\hat{x}(t)$ are their respective derivatives, they are also completely characterized by the samples and match exactly. So our reconstruction using this multi-channel time decoding algorithm is perfect in the noiseless case.

    Our findings are summarized into the following theorem.
    \begin{thm}
    	\label{thm: multi TEM converges}
		Assume $x(t)$ is a $2\Omega$-bandlimited signal in $L^2(\mathbb{R})$ that is bounded such that $|x(t)|\leq c$ and that has a well-defined integral $\int_{-\infty}^t x(u)\, du = \gamma <\infty$. If $x(t)$ is passed through $M$ TEMs with parameters $\kappa$, $\delta$ and $b$, such that $b>c$, the shifts $\alpha_i$, $i=1\cdots M$ between the TEMs are nonzero and \[\Omega<\frac{M\pi(b-c)}{2\kappa\delta},\] then 
		\begin{equation}
		\lim_{\ell \rightarrow \infty} x_\ell(t) = x(t).
		\end{equation}
		when $x_\ell(t)$ is as defined in~\eqref{eq: Rec operator single channel}-\eqref{eq: Rec Alg M channel inductive step},
    \end{thm}

    We have thus shown that using $M$ time encoding machines to encode a $2\Omega$-bandlimited signal $x(t)$ allows a bandwidth which is $M$ times larger than in the single channel case, no matter how the shifts between the machines are configured, as long as they are all nonzero. As already stated, we had shown in~\cite{adam2019multi-channel} that the bandwidth could become $M$ times larger, but only if the machines were configured in such a way that their integrators had equally spaced values. In other words, if we denote $y_{i}(t)$ to be the value of the integrator of machine $i$, then we required, $\forall i=1\cdots M-1$,
    \begin{equation}
        \alpha_i = y_{i+1}(t)-y_i(t) = 2\delta/M.
    \end{equation}
    Configuring the integrator shifts between two machines is not easy (somewhat like synchronizing the clocks of different channels in classical sampling). Therefore, achieving maximal information gain becomes a harder feat. Here, we have shown that the bandwidth improvement by a factor of $M$ is actually independent of the $\alpha_i$'s, as long as these are nonzero, and that the reconstruction algorithm does not require the knowledge of the $\alpha_i$'s.

    The strength of this algorithm lies in its simplicity. We have $M$ TEMs with integrators that are shifted with respect to each other by some shifts $\alpha_i$, and if the set of $\alpha_i$'s changes, the spike outputs of the machines change. However, this algorithm does not require knowledge of the shifts, it only operates on the spike times generated by the machine. Moreover, labelling of spike times according to the machine they come from is not necessary. TEMs are shifted with respect to each other by $\alpha_i$, so the order of spiking of the machines is fixed: we will always have spikes coming from TEM $\mathrm{A}_1,\mathrm{A}_2,\cdots,\mathrm{A}_M,\mathrm{A}_1,\mathrm{A}_2,\cdots$. Therefore, the algorithm operates on a model as depicted in Fig.~\ref{fig: M-Channel TEM}, and is still able to disentangle spike streams.

\subsection{Closed Form Solution}

    We have described a POCS iterative algorithm to reconstruct an input signal from the output of a TEM. However, adopting a POCS algorithm in practice might be quite slow, and the performance is dependent on the number of iterations or the stopping criteria. Instead, for practical implementations, we propose an equivalent closed form solution to the problem.

    First, let $\left\{\tilde{t}_k,k\in\mathbb{Z}\right\}$ denote the set of combined and ordered spike times from all machines A$_1$, $\cdots$, A$_M$. Now define
    \begin{equation}
        \tilde{\mathcal{G}}\left(y\right) =   \sum_{k \in \mathbb{Z}}y_k \tilde{g}_{[t_k, t_{k+1})}(t),
    \end{equation}
    where $\tilde{g}_{[t_k, t_{k+1})}(t) = \mathbbm{1}_{[t_k, t_{k+1})}(t)\ast g(t)$.

    Also define \[\mathbf{\tilde{q}} = \left[\int_{\tilde{t}_k}^{\tilde{t}_{k+M}}x(u)\, du\right]_{k \in \mathbb{Z}},\] and \[\mathbf{\tilde{H}}= {\left[\tilde{H}_{\ell k}\right]_{\ell,k \in \mathbb{Z}} = \left[\int_{\tilde{t}_\ell}^{\tilde{t}_{\ell+M}}\tilde{g}_{[t_k, t_{k+1})}(u)\, du \right]_{\ell,k \in \mathbb{Z}}}.\]
    Then, one can show by induction that $x_\ell$, as defined in~\eqref{eq: Rec operator M channel}~-~\eqref{eq: Rec Alg M channel inductive step}, can be expressed as
    
    \begin{equation}
	    x_\ell = \tilde{\mathcal{G}} \left( \sum_{k=0}^\ell \left(\mathbf{I}-\mathbf{\tilde{H}}\right)^k \mathbf{\tilde{q}}\right).
	\end{equation}
    
    Now we note that $\lim_{\ell\rightarrow \infty} \sum_{k=0}^\ell \left(\mathbf{I}-\mathbf{\tilde{H}}\right)^k = \mathbf{\tilde{H}^{+}}$, where$\mathbf{\ ^+}$ denotes a pseudo-inverse. Therefore, a closed form solution for reconstructing $x(t)$ under the same conditions as the ones posed in Theorem~\ref{thm: multi TEM converges} is
    \begin{equation}
    \label{eq: Multi-Channel Closed Form Reconstruction}
        \hat{x}(t) = \tilde{\mathcal{G}}\left( \mathbf{\tilde{H}}^+ \mathbf{\tilde{q}}\right).
    \end{equation}

\section{Simulations}
\label{sec: convergence speed}

	\subsection{Simulation Setup}
	To validate our theory, we set up an environment to test our reconstruction algorithm, and verify its performance under different conditions.
	Generally, our approach includes four steps:
	\begin{enumerate}
	    \item \textit{Signal generation:} We wish to generate signals that are $2\Omega$-bandlimited, and sample them over a finite time window $[t_{start}, t_{end}]$. We assume the signals are a linear combination of sincs centered at uniform time points  between $t_{start}$ and $t_{end}$ with a separation of $\pi/\Omega$. The amplitudes of these sincs are randomly generated assuming a uniform distribution over $[0,1]$, and the signals are finally all normalized to have unit norm.
	    \item \textit{Signal sampling:} Time encoding is performed using either of two techniques. In the first, we sample the signal discretely using small steps and approximate the integral of the signal with a cumulative sum. In the second technique, signals are assumed to be generated according to our signal generation procedure described above, and we search for each spike time using binary search, by evaluating  the signal's integral at different time points. On one hand, the first technique is more versatile to different signal types. On the other hand, the second technique allows more spike time precision without requiring extra space requirements which arise from heavily oversampling the input signal.
	    \item \textit{Signal reconstruction:} Signal reconstruction is performed using the closed form solution provided in~\eqref{eq: Multi-Channel Closed Form Reconstruction}. The reconstruction can also be done using the iterative POCS algorithm, but obtaining the reconstruction then becomes more time consuming and the reconstruction's performance depends on the chosen number of iterations or the stopping criterion.
	    \item \textit{Performance evaluation:} To evaluate the performance of our reconstruction, we compute the difference between our (discretized) reconstruction and the original (discretized) signal. We then compute the power of this difference for the middle $90\%$ of the signal (assuming the start and end generally have a less precise reconstruction because of our finite support sampling and reconstruction setup). We call this the mean-squared reconstruction error. As all signals are normalized to have unit norm, the mean-squared reconstruction error of different signals are comparable.
	\end{enumerate}
	
		\begin{figure}[tb]
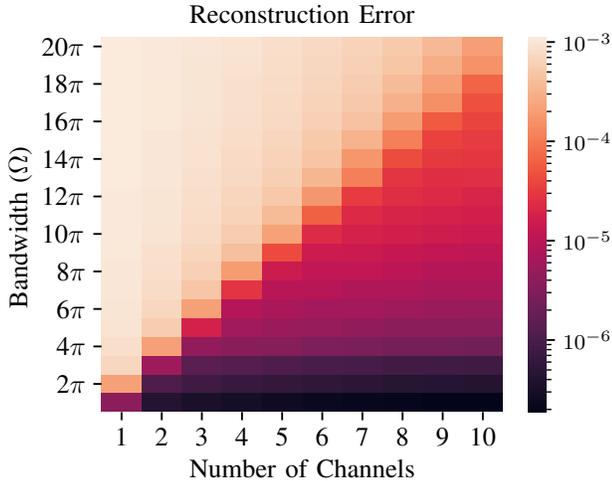

	\begin{minipage}[b]{1.0\linewidth}
		\centering
		\centerline{		        
		\def\svgwidth{\columnwidth}
		\subfile{figures/Figure8_VarNumChannels.tex}}
	\end{minipage}
	\caption{Error of time encoding reconstruction when $M = 1\cdots 10$ channels with equally-shifted integrators encode a signal as its bandwidth varies. The mean-squared error is averaged over one hundred trials and plotted as a function of the bandwidth and the number of samples.}
	\label{fig: Simulating Rec}
	\end{figure}
	
	In the simulations that come, we will evaluate the reconstruction algorithm's success while varying 4 main variables in different combinations: the bandwidth of the sampled signals $\Omega$, the number of machines $M$, the shifts $\alpha_i$ between the machines and the variance of the noise added on top of the spike times.
	
	The parameters of the TEMs are always kept constant, taking $\kappa=\delta=1$ and $b=\max_t|x(t)|+1$ where $x(t)$ is the input signal. 
	In fact, when evaluating the algorithm performance, we can keep these parameters constant without loss of generality as long as we vary $\Omega$. Indeed, assume we have a $2\Omega$-bandlimited signal $x_1(t)$ that is sampled using parameters $\kappa$, $\delta$ and $b$ and generates spike times $\left\lbrace t_k(x_1), k \in \mathbb{Z}\right\rbrace$. Then, let $x_2(t)$ be a $2p\Omega$-bandlimited signal such that $x_2(t) = x_1(pt)$. Now assume $x_2(t)$ is sampled using parameters $\kappa$, $\delta/p$ and $b$ and generates the spike times $\left\lbrace t_k(x_2), k \in \mathbb{Z}\right\rbrace$, then $t_{k+1}(x_2)-t_{k}(x_2) = (t_{k+1}(x)-t_k(x))/p$. In essence, the information content of the two signals is the same, and the increase in bandwidth of one can be compensated for by a decrease in the threshold $\delta$ and vice versa. One can also perform similar analyses for the other parameters $\kappa$ and $b$. Therefore we decide to fix the first three of the four parameters $\kappa$, $\delta$, $b$ and $\Omega$ and only vary the last one.

	The figures are reproducible using code available online~\cite{karen_adam_2019_3558507}. 
	
    \subsection{Experimental Validation of Theorem 1}
    In Fig.~\ref{fig: Simulating Rec}, we randomly generate one hundred $2\Omega$-bandlimited signals, for each value of $\Omega = \pi, 2\pi, \cdots, 20\pi$. We provide the reconstruction error when using $M = 1\cdots 10$ channels with the same parameters $\kappa$, $\delta$ and $b$ to sample and reconstruct the signals. The channels are constructed with equally spaced shifts, i.e. for an $M$-channel TEM, the integrator shifts are $\alpha_i=2\delta/M, \ \forall i = 1\cdots M$. 
    For every number of channels $M$ used to perform the signal sampling and reconstruction, we have a different constraint on the bandwidth which ensures that this $M$-channel TEM can reconstruct its input signal as given in~\eqref{eq: Omega bound multi}. Looking at Fig.~\ref{fig: Simulating Rec}, for each number of channels $M$, we can see a degradation of the reconstruction as the bandwidth increases beyond the constraint placed in~\eqref{eq: Omega bound multi}. Notice how this degradation happens for higher $\Omega$ as the number of channels $M$ increases. The separation between ``good" and ``bad" performance seems to change linearly with the number of channels $M$ as we would expect from~\eqref{eq: Omega bound multi}.

    \begin{figure}[tb]
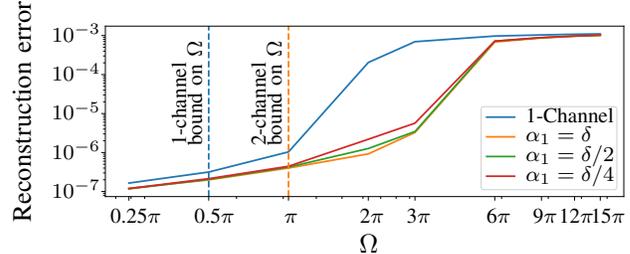

	\begin{minipage}[b]{1.0\linewidth}
		\centering
		\centerline{
		\def\svgwidth{\columnwidth}
		\subfile{figures/Figure9.tex}}
	\end{minipage}
	\caption{Error of time encoding reconstruction, when using a single channel (blue), and when using 2 channels with different spacing configurations (orange, green, red). The mean-squared error is averaged over two hundred trials and plotted as a function of the bandwidth. The shift takes value $\alpha_1$, i.e. TEM A$_2$ has its integrator $\alpha$ ahead of TEM A$_1$ (modulo $2\delta$) and consequently, TEM A$_1$ has its integrator $2\delta-\alpha$ ahead of TEM A$_2$ (modulo $2\delta$). The closer the shift is to $\delta$, the more equally spaced the samples are expected to be.}
	\label{fig: Independence of shift}
	\end{figure}
    
    To show that this $M$-fold improvement on the bound for the bandwidth is independent of the value of the shift, we evaluate the reconstruction error when signals are sampled using 2-channel TEMs with different values for the shift. In Fig.~\ref{fig: Independence of shift}, we again simulate one hundred $2\Omega$-bandlimited signals where $\Omega$ now varies between $\pi/4$ and $15\pi$, and plot the averaged reconstruction error for 2-channel decoding, as well as the averaged reconstruction error for single-channel decoding. 
    For both the single-channel and the 2-channel case, the reconstruction error is low for low values of $\Omega$ and becomes much higher as $\Omega$ surpasses the bound provided in~\eqref{eq: Omega bound multi} and plotted using the dashed vertical lines in Fig.~\ref{fig: Independence of shift}. Notice how the reconstruction is successful for wider ranges of the bandwidth in the 2-channel case, compared to the single-channel case, and how different values of the integrator shifts between the two channels do not affect this region of success.

	\begin{figure}[tb]
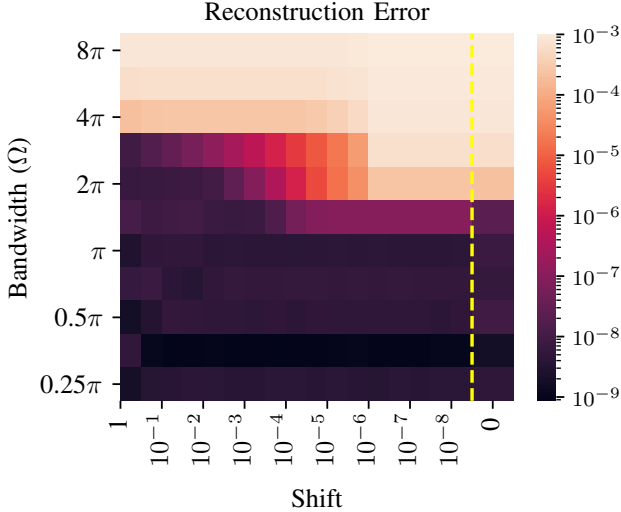

	\begin{minipage}[b]{1.0\linewidth}
		\centering
		\centerline{
		\def\svgwidth{\columnwidth}
		\subfile{figures/Figure10_VarShifts.tex}}
	\end{minipage}
	\caption{Reconstruction error plotted as a function of bandwidth and integrator shift. A hundred $2\Omega$-bandlimited signals, where $\Omega$ varies between $\pi/4$ and $8\pi$ are generated and subsequently sampled and reconstructed using two-channel time encoding and decoding. The TEM used has fixed parameters $\kappa=1$, $\delta=1$ and $b=\max_t|x(t)|+1$  but variable integrator shifts. Here, we plot one of the two shifts $\alpha_1$, the value of the second shift can be obtained from $\alpha_2 = 2\delta - \alpha_1$. The mean-squared error is averaged over the hundred randomly generated signals and plotted as a function of bandwidth and shift. Although we have shown that the value of the integrator shifts should have no effect on the reconstructible bandwidth~\eqref{eq: Omega bound multi}, very small shifts perform  less well than shifts that are within the same order of magnitude as the threshold $\delta$. The rightmost column, separated by the dashed yellow line, shows the reconstruction error when a shift of zero is used. In other words, this is the reconstruction error when using single-channel time encoding and decoding.}
	\label{fig: vary shift and omega}
	\end{figure}

	\subsection{Problem Ill-Conditioning for Small Shifts}
	We have shown that, in theory, the condition we placed in~\eqref{eq: Omega bound multi} is sufficient for the reconstruction algorithm to converge no matter the shifts between the integrators of different machines. Moreover, Fig.~\ref{fig: Independence of shift} verified this result for a few values of the integrator shifts. Intuitively, however, the problem should become more ill-posed as the shifts approach zero.
	
	To investigate this, we evaluate the performance of two-channel time encoding and decoding as the shifts between the channels approach zero. We randomly generate one hundred $2\Omega$-bandlimited signals, where $\Omega$ varies between $\pi/4$ and $8\pi$. These signals are then encoded and decoded using two-channel TEMs with fixed parameters $\kappa$, $\delta$ and $b$ and with varying shift $\alpha$. We then estimate the reconstruction success by computing the reconstruction error.

	Figure~\ref{fig: vary shift and omega} is essentially a two dimensional version of the plot in Fig.~\ref{fig: Independence of shift} which investigates smaller shifts. As the integrator shift approaches zero, the outputs of the two channels of the TEM start to resemble each other more and more, so our two-channel encoding starts to resemble single-channel encoding. Therefore, we also include, in Fig.~\ref{fig: vary shift and omega} the reconstruction error of a single-channel TEM, to compare it to the result obtained with two-channel encoding with very small shift.
	
	Note that the system seems to perform reasonably well in the noiseless case, when the condition in~\eqref{eq: Omega bound multi} is satisfied, for values of the shift that are not too small (less than $10^{-4}$). 
	
	\subsection{Algorithm Performance in Noisy Settings}
	
    \begin{figure}[tb]
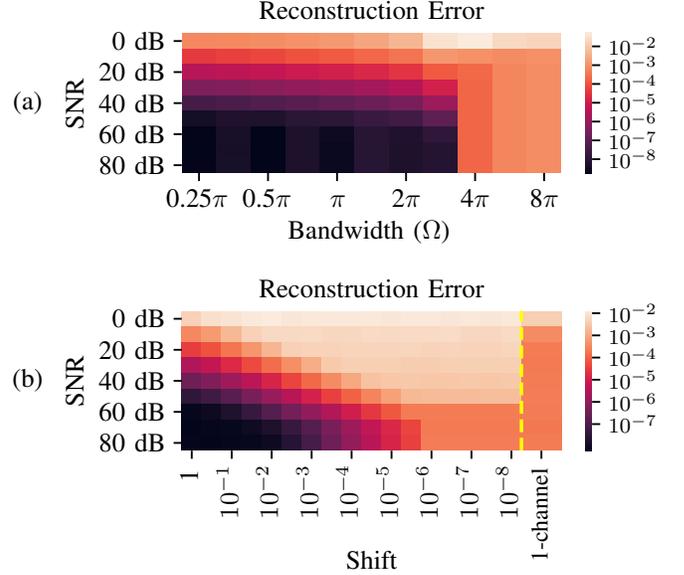

	\begin{minipage}[b]{1.0\linewidth}
		\centering
		\centerline{
		\def\svgwidth{\columnwidth}
		\subfile{figures/Figure11_combined.tex}}
    \vspace{-20pt}
	\end{minipage}
	\caption{(a) Reconstruction error of two-channel time encoding with equal spacing in the shifts, with gaussian noise added to the spike times, when the SNR varies between 80 dB and 0 dB and the bandwidth varies between $0.25\pi$ and $8\pi$.\\
    (b) To the left of the yellow line, reconstruction error of two-channel time encoding of $2\Omega$-bandlimited signals with $\Omega = 2\pi$, with gaussian noise added to the spike times, when the SNR varies between 80 dB and 0 dB and the integrator shifts vary between 1 (equal spacing) and $10^{-8}$. To the right of the yellow line, reconstruction error of single-channel time encoding of signals with the same bandwidth.}
	\label{fig: noise}
	\end{figure}
	
	We now provide basic analyses to understand the system's performance in the case of noise. We study the effect of noise on reconstruction when varying two other parameters: the bandwidth (Fig.~\ref{fig: noise}.a) and the shift (Fig.~\ref{fig: noise}.b). In both scenarios, we assume that we have a two-channel TEM with parameters $\kappa$, $\delta$ and $b$ fixed, such that the TEM is guaranteed to be able to reconstruct $2\Omega$-bandlimited signals with $\Omega = \pi$.
	
	In Fig.~\ref{fig: noise}.a, we assume that the two machines are equally spaced ($\alpha_1 = \alpha_2$) and that Gaussian noise is added to the spike times. We then vary the SNR between 80 dB and 0 dB and the bandwidth between $0.25\pi$ and $8\pi$. Notice how, with high SNR, i.e. with low noise, the machines can reconstruct signals with bandwidths that go up to $2\pi$. As the SNR becomes lower, the machines become less apt at reconstructing signals with high bandwidths.
	
	In Fig.~\ref{fig: noise}.b, we assume that the input signals to the machines are $2\Omega$-bandlimited where $\Omega = 2\pi$. We then vary the SNR between 80 dB and 0 dB and the integrator shifts of the machines between $1$ (equal spacing) and $10^{-8}$. Notice how the reconstruction of the input becomes worse as the shift between the machines approaches zero and as the SNR decreases. We also provide the reconstruction error for the single-channel time encoding as a comparison.

	\subsection{Setting Shift Values: Practical Considerations}

	As seen in Fig.~\ref{fig: vary shift and omega}, the algorithm we have suggested works well, in the noiseless case, for varying values of the shift, until the shift approaches zero very closely. Therefore, if the shifts are randomly assigned, they will, with very high probability fall in a regime where the algorithm provides good performance.
	
	As for the physical implementation of the shifts, as previously mentioned, the hardware implementation of time encoding machines uses a capacitor which can hold some initial charge. One can make sure that different machines have different initial charges on their capacitors by first feeding the TEMs with different signals to randomly initialize the values of the capacitor before beginning to encode the signal of interest.

\section{Conclusion}
We have studied multi-channel time encoding of $2\Omega$-bandlimited signals, proposed an algorithm for reconstructing an input signal from its samples, and provided sufficient conditions on $\Omega$ for the algorithm to converge to the correct solution. We have shown that if a TEM can perfectly encode a $2\Omega$-bandlimited signal, then $M$ TEMs with the same parameters and with shifts in their integrators can perfectly encode a $2M\Omega$-bandlimited signal. The reconstruction algorithm is then based on a projection onto convex sets method.

The improvement on bandwidth that we found is independent of the value of the shifts between the machine integrators, as long as these shifts are nonzero. We have also shown that the knowledge of the relative shifts between the machines is not necessary for reconstruction to be possible. This is not the case in similar setups of multi-channel encoding in the classical sampling scenario where an unknown shift makes the inverse problem more difficult to solve.

Our setting has focused on reconstructing signals using TEMs with the same parameters $\kappa$, $\delta$ and $b$, where $b>0$. However, our algorithm can be extended to scenarios where these parameters vary according to the machine, and to scenarios where $b=0$. However, the improvement in bandwidth becomes harder to quantify, as different configurations entail different improvements and many edge cases can be found. As a general rule, a $2\Omega$-bandlimited signal can still be reconstructed, if $\Omega$ is inversely proportional to the number of linearly independent constraints that arise from the spike times generated by the machines.

In our future work, we address this kind of setup and extend it to scenarios with vectors of input that are fed into the TEMs with different combinations~\cite{adam2019multi-channel}. We also hope to understand what potential sources of noise are in practice, and study our setup's behavior in noisy conditions.

\begin{appendices}
\section{Previous results}\label{sec: Appendix Former Results}

The following results can all be found in~\cite{feichtinger1994theory}.

\begin{lemma}[Bernstein's inequality]
\label{lemma: bernstein's inequality}
		If $x=x(t)$ is a function defined on $\mathbb{R}$ bandlimited to $[-\Omega, \Omega]$ then $dx/du$ is also bandlimited and 
		\begin{equation*}
		\left\lVert{\frac{dx}{du}}\right\rVert^2 \leq \Omega \norm{x}^2.
		\end{equation*}
\end{lemma}
	
\begin{lemma}[Wirtinger's inequality]
\label{lemma: wirtinger's inequality}
		If $x,dx/dt \in L^2(a,b)$ and either $x(a) = 0$ or $x(b)=0$, then\footnote{A signal $x(t)$ is in $L^2(a,b)$ if $\lVert x(t) \rVert_2 = \left(\int_a^b|x(u)|^2\, du\right)^{1/2} < \infty$.}
		\begin{equation*}
		\int_{a}^{b} \left|x(u)\right|^2du \leq \frac{4}{\pi^2}(b-a)^2\int_{a}^{b}\left|\frac{dx}{du}\right|^2du.
		\end{equation*}
\end{lemma}

    \begin{definition}\label{def: relatively separated}
    A set $\left\{t_k, k \in \mathbb{Z}\right\}$ is called \textit{relatively separated} if it can be divided into a finite number of subsets so that $\left|x_n-x_m\right|\geq \beta>0$ for a fixed $\beta>0$, $n\neq m$, and $x_n$, $x_m$ in the same subset.
    \end{definition}

\section{Proofs}\label{sec: Appendix Proofs}
    \begin{manualtheorem}{\ref{lemma: P2 projection operator}}
    $\mathcal{P}_{\Omega}$ is a firmly nonexpansive projection operator onto $\mathcal{C}_{\Omega}$.
\end{manualtheorem}

\begin{proof}[Proof of Lemma~\ref{lemma: P2 projection operator}]
    First, we show that $\mathcal{P}_{\Omega}$ is idempotent. Denoting $G(\omega)$ to be the Fourier transform of $g(t)$, we get $G(\omega)=1, \, \forall |\omega| \leq \Omega$ and zero otherwise.
    
    Then,
    \begin{equation*}
    \mathcal{P}_{\Omega}\left(\mathcal{P}_{\Omega}\left(y(t)\right)\right) = y(t)\ast g(t)\ast g(t) = y(t) \ast g(t),
    \end{equation*}
    since $G(\omega)^2 = G(\omega)$, so that $g(t)\ast g(t) = g(t)$.
    
    We now show that $\mathcal{P}_{\Omega}$ has as range the space $\mathcal{C}_\Omega$ of $2\Omega$-bandlimited functions in $L^2(\mathbb{R})$. 
    
    First, let $y(t)$ be an arbitrary $L^2(\mathbb{R})$ function, its Fourier transform $Y(\omega)$ is then also $L^2(\mathbb{R})$, according to Parseval's theorem. The result of the projection will have Fourier transform $Y(\omega)G(\omega) = Y(\omega), \, \forall\,  |\omega|<\Omega$, and zero otherwise. Therefore $Y(\omega)G(\omega)$ is also $\in L^2(\mathbb{R})$, in addition to it being in $\mathcal{P}_{\Omega}$.
    
    Now, let $y(t)$ be a $2\Omega$-bandlimited function $\in L^2(\mathbb{R})$, then its Fourier transform $Y(\omega)$ is such that $Y(\omega) = 0, \forall |\omega|>\Omega$. Convolving $y(t)$ with $g(t)$ in the time domain only multiplies $Y(\omega)$ by $1$ in the region where it is nonzero. Therefore $y(t)\ast g(t) = y(t) \in \mathcal{C}_\Omega$.
    
    Finally, $\mathcal{P}_{\Omega}$ is firmly non-expansive, since it is an orthogonal projection operator~\cite{vetterli2014foundations}.
\end{proof}
 \vspace{15pt}
 
 \begin{manualtheorem}{\ref{lemma: C2 convex}}
$\mathcal{C}_\Omega$ is convex.
\end{manualtheorem}

\begin{proof}[Proof of Lemma~\ref{lemma: C2 convex}]
Let $y_1(t)$ and $y_2(t)$ be in $\mathcal{C}_{\Omega}$. Then let $y_3(t)$ be any convex combination of $y_1(t)$ and $y_2(t)$, i.e. $y_3(t) = \lambda y_1(t) + (1-\lambda)y_2(t)$, where $\lambda \in [0,1]$. Then, let $Y_1(\omega)$, $Y_2(\omega)$ and $Y_3(\omega)$ be the Fourier transforms of $y_1(t)$, $y_2(t)$ and $y_3(t)$ respectively. By linearity of the Fourier transform, we find that $Y_3(\omega) = \lambda Y_1(\omega)+(1-\lambda)Y_2(\omega), \ \forall \omega\in\mathbb{R}$. Therefore, since $y_1(t)$ and $y_2(t)$ are in $\mathcal{C}_{\Omega}$ and $Y_1(\omega)=Y_2(\omega)=0\ \forall |\omega|>\Omega$, $Y_3(\omega)=0\ \forall |\omega|>\Omega$. Therefore, $y_3(t)$ is also $2\Omega$-bandlimited. $L^2(\mathbb{R})$ is also a convex set (as it is a linear space), therefore $y_3(t)$ is also in $L^2(\mathbb{R})$, as $y_1(t)$ and $y_2(t) \in L^2(\mathbb{R})$. Therefore $y_3(t) \in \mathcal{C}_{\Omega}$, thus showing that $\mathcal{C}_\Omega$ is convex.
\end{proof}

\vspace{15pt}

 \begin{manualtheorem}{\ref{lemma: P1 projection operator}}
    $\mathcal{P}_{\mathrm{A}_1}$ is a firmly nonexpansive projection operator onto $\mathcal{C}_{\mathrm{A}_1}$.

\end{manualtheorem}

\begin{proof}[Proof of Lemma~\ref{lemma: P1 projection operator}]
    First we show that $\mathcal{P}_{\mathrm{A}_1}$ is idempotent. Note that $\int_{t_k}^{t_{k+1}} \mathcal{P}_{\mathrm{A}_1}y(u) = \int_{t_k}^{t_{k+1}} x(u) \, du, \ \forall k \in \mathbb{Z}$. Therefore,
    \begin{align}
        \mathcal{P}_{\mathrm{A}_1}&\left(\mathcal{P}_{\mathrm{A}_1}\left(y(t)\right)\right) \notag\\
        &= \mathcal{P}_{\mathrm{A}_1}\left(y(t)\right) \notag\\
        &\phantom{{}=} + \sum_{k \in \mathbb{Z}} \int_{t_k}^{t_{k+1}} \left[x(u)-\mathcal{P}_{\mathrm{A}_1}\left(y(u)\right)\right] \, du \  \frac{\mathbbm{1}_{[t_k, t_{k+1})}(t)}{t_{k+1}-t_k}\ \notag\\
        & = \mathcal{P}_{\mathrm{A}_1}\left(y(t)\right). \notag
    \end{align}
    Now we show that the range of $\mathcal{P}_{\mathrm{A}_1}$ is indeed the space of functions $z(t)$ with  $\int_{t_k}^{t_{k+1}}z(u)\, du = \int_{t_k}^{t_{k+1}}x(u)\, du$. 
    
    First let $y(t)$ be a function in $L^2(\mathbb(R)$. It is easy to show that $\mathcal{P}_{\mathrm{A}_1}\left(y(t)\right)$ will be have $\int_{t_k}^{t_{k+1}}\mathcal{P}_{\mathrm{A}_1}\left(y(u)\right)\, du = \int_{t_k}^{t_{k+1}}x(u)\, du$. One can also show that $\mathcal{P}_{\mathrm{A}_1}\left(y(t)\right)$ will be in $L^2(\mathbb{R})$, by using Lemma~\ref{lemma: B1 projects on L2}.

    Now,  let $y(t)$ be in $\mathcal{C}_{\mathrm{A}_1}$, then
    \begin{align}
        \mathcal{P}_{\mathrm{A}_1}\left(y(t)\right) &= y(t) \notag\\
        &\phantom{{}=} + \sum_{k \in \mathbb{Z}} \int_{t_k}^{t_{k+1}} \left[x(u)-y(u)\right] \, du \ \ \frac{\mathbbm{1}_{[t_k, t_{k+1})}(t)}{t_{k+1}-t_k}\notag\\
        &= y(t) + \sum_{k \in \mathbb{Z}} 0 \times \frac{\mathbbm{1}_{[t_k, t_{k+1})}(t)}{t_{k+1}-t_k}\notag\\
        &=y_1(t).\notag
    \end{align}
    Therefore, $\mathcal{P}_{\mathrm{A}_1}$ has range $\mathcal{C}_{\mathrm{A}_1}$.
    
    It remains to show that $\mathcal{P}_{\mathrm{A}_1}$ is firmly nonexpansive. To do so, it is sufficient to show that  $\mathcal{P}_{\mathrm{A}_1}$ can be written
		\begin{equation*}
		\mathcal{P}_{\mathrm{A}_1} = \frac{1}{2}\mathcal{I}+\frac{1}{2}\mathcal{N},
		\end{equation*}
		where $\mathcal{N}$ is an nonexpansive operator.
		Indeed, the operator can be written as such if we set
		\begin{equation*}
		\mathcal{N}y(t) = \mathcal{I}y(t) + 2\mathcal{B}_1\left( x(t) - y(t)\right).
		\end{equation*}
		We want to show that $\mathcal{N}$ is nonexpansive, therefore it is sufficient to show that for any $y_1(t)$ and $y_2(t)$ in $L^2(\mathbb{R})$,
		\begin{equation*}
		\left| \left|\mathcal{N}y_1 - \mathcal{N}y_2 \right| \right| \leq \left| \left|y_1 - y_2 \right| \right|.
		\end{equation*}
		We will start with the left hand side of the equation:
		\begin{align}
		\left| \left|\mathcal{N}y_1 - \mathcal{N}y_2 \right| \right| &= \left| \left| \mathcal{I}y_1(t) + 2\mathcal{B}_1\left( x(t) - y_1(t)\right) \right. \right.\notag\\
		&\phantom{{}=} - \left.\left.\mathcal{I}y_2(t) - 2\mathcal{B}_1\left( x(t) - y_2(t)\right) \right| \right|\notag\\
		&= \left| \left| \mathcal{I}y_1(t) - 2\mathcal{B}_1y_1(t) - \mathcal{I}y_2(t) + 2\mathcal{B}_1 y_2(t) \right| \right|\notag\\
		&= \left| \left| \left(\mathcal{I} - 2\mathcal{B}_1\right)\left(y_1-y_2\right) \right| \right|\notag\\
		&\leq \left| \left| \mathcal{I} - 2\mathcal{B}_1\right| \right| \left| \left| y_1-y_2 \right| \right| \notag\\
		&\leq \left| \left| y_1-y_2 \right| \right| \notag
		\end{align}
		as we have shown, in Lemma~\ref{lemma: Op norm I-2B =1} that $\norm{\mathcal{I} - 2\mathcal{B}_1}=1$.
\end{proof}

\vspace{15pt}
\begin{manualtheorem}{\ref{lemma: C1 convex}}
$\mathcal{C}_{\mathrm{A}_1}$ is convex.
\end{manualtheorem}
\begin{proof}[Proof of Lemma~\ref{lemma: C1 convex}]
Let $y_1(t)$ and $y_2(t)$ be in $\mathcal{C}_{\mathrm{A}_1}$. Then let $y_3(t)$ be any convex combination of $y_1(t)$ and $y_2(t)$, i.e. $y_3(t) = \lambda y_1(t) + (1-\lambda)y_2(t)$, where $\lambda \in [0,1]$. Then, we have:
\begin{align}
    \int_{t_k}^{t_{k+1}}y_3(t)\, dt &= \int_{t_k}^{t_{k+1}} \lambda y_1(t) + (1-\lambda)y_2(t)\, dt\notag\\
     &= \lambda\int_{t_k}^{t_{k+1}}  y_1(t) \, dt+ (1-\lambda)\int_{t_k}^{t_{k+1}}y_2(t)\, dt\notag\\
    &= \lambda\int_{t_k}^{t_{k+1}} x(t) \, dt+ (1-\lambda)\int_{t_k}^{t_{k+1}}x(t)\, dt\notag\\
    &=\int_{t_k}^{t_{k+1}}x(t) \, dt. \notag
\end{align}
The first equality holds because of the definition of $y_3(t)$, and the third  equality holds because  $y_1(t)$ and $y_2(t)$ are in $\mathcal{C}_{\mathrm{A}_1}$. The result shows that $y_3(t)$ is also consistent with the spike times $\left\lbrace t_k, k \in \mathbb{Z}\right\rbrace$. On the other hand, $L^2(\mathbb{R})$ is a linear space (and therefore a convex set), so $y_3(t) \in L^2(\mathbb{R})$ as well. Therefore, $y_3(t)\in \mathcal{C}_{\mathrm{A}_1}$, thus proving that $\mathcal{C}_{\mathrm{A}_1}$ is a convex set.
\end{proof}

\vspace{15pt}
    \begin{lemma}
		\label{lemma: B1 projects on L2}
		If $y(t)$ is in $L^2(\mathbb{R})$, then $\mathcal{B}_1y(t)$ is also in $L^2(\mathbb{R})$.
	\end{lemma}
	\begin{proof}
		Let $y(t) \in L^2(\mathbb{R})$, so $\int_{-\infty}^{\infty}\left|y(t)\right|^2 = d <\infty$ for some $d\in\mathbb{R}$.
		\begin{align}
		\int_{-\infty}^{\infty} \left|\mathcal{B}_1y(t)\right|^2 \, dt &\stackrel{(a)}{=} 	\sum_{k\in\mathbb{Z}}\int_{t_k}^{t_{k+1}} \left|\mathcal{B}_1y(t)\right|^2 \, dt \notag\\
		&\stackrel{(b)}{=} 	\sum_{k\in\mathbb{Z}}\int_{t_k}^{t_{k+1}} \left|\frac{\int_{t_k}^{t_{k+1}}y(u)\, du}{t_{k+1}-t_k} \right|^2 \, dt\notag\\
        &\stackrel{(c)}{=} \sum_{k\in\mathbb{Z}}\left(t_{k+1}-t_k\right) \left|\frac{\int_{t_k}^{t_{k+1}}y(u)\, du}{t_{k+1}-t_k} \right|^2 \, dt\notag
		\end{align}
		\begin{align}
		\int_{-\infty}^{\infty} \left|\mathcal{B}_1y(t)\right|^2 \, dt &\stackrel{(d)}{=} \sum_{k\in\mathbb{Z}} \frac{\left(\int_{t_k}^{t_{k+1}}y(u)\, du\right)^2}{t_{k+1}-t_k}\, dt\notag\\
		&\stackrel{(e)}{\leq} \sum_{k\in\mathbb{Z}} \left(t_{k+1}-t_k\right)\frac{\int_{t_k}^{t_{k+1}}\left(y(u)\right)^2\, du}{t_{k+1}-t_k}\, dt, \notag\\
		&\stackrel{(f)}{=} \sum_{k\in\mathbb{Z}} \int_{t_k}^{t_{k+1}}\left(y(u)\right)^2\, du\, dt\notag\\
		&\stackrel{(g)}{=} \int_{-\infty}^{\infty}\left(y(u)\right)^2\, du\, dt = d <\infty.\notag
		\end{align}
		Here, inequality (e) arises from the Cauchy-Schwarz inequality.
	\end{proof}
	
	\vspace{15pt}
	
\begin{lemma}
\label{lemma: Op norm I-2B =1}
The operator norm of operator $\mathcal{I}-2\mathcal{B}_1$ is $\left| \left| \mathcal{I}-2\mathcal{B}_1 \right| \right| = 1$.
\end{lemma}
\begin{proof}
	To find the operator norm $\left| \left| \mathcal{I}-2\mathcal{B}_1 \right| \right| = \sup_y \frac{\norm{(\mathcal{I}-2\mathcal{B}_1)y}}{\norm{y}}$, let us compute
	\begin{align}
	\left| \left| \left(\mathcal{I} - 2\mathcal{B}_1\right) y(t) \right|\right|^2 &= \int_{-\infty}^{\infty} \left|y(t) -2\mathcal{B}_1y(t)\right|^2 dt \notag\\
	& = \sum_{k \in \mathbb{Z}}\int_{t_k}^{t_{k+1}} \left(y(t) -2\mathcal{B}_1y(t)\right)^2 dt \notag\\
	& = \sum_{k \in \mathbb{Z}}\int_{t_k}^{t_{k+1}} \left(y(t) -2\frac{\int_{t_k}^{t_{k+1}}y(u)\, du}{t_{k+1}-t_k}\right)^2 dt \notag\\
	& = \sum_{k \in \mathbb{Z}}\int_{t_k}^{t_{k+1}} \left(y(t)\right)^2 +4\left(\frac{\int_{t_k}^{t_{k+1}}y(u)\, du}{t_{k+1}-t_k}\right)^2 \notag\\
	&\quad - 4\frac{\int_{t_k}^{t_{k+1}}y(u)\, du}{t_{k+1}-t_k}x(t) \, dt \notag\\
    & = \sum_{k \in \mathbb{Z}} 4\left(t_{k+1}-t_k\right)\left(\frac{\int_{t_k}^{t_{k+1}}y(u)\, du}{t_{k+1}-t_k}\right)^2 \notag\\
	& \quad - 4\frac{\left(\int_{t_k}^{t_{k+1}}y(u)\, du\right)^2}{t_{k+1}-t_k} + \int_{t_k}^{t_{k+1}} \left(y(t)\right)^2  \, dt \notag\\
	& = \sum_{k \in \mathbb{Z}} 4\frac{\left(\int_{t_k}^{t_{k+1}}y(u)\, du\right)^2}{t_{k+1}-t_k} \notag\\
	&\quad - 4\frac{\left(\int_{t_k}^{t_{k+1}}y(u)\, du\right)^2}{t_{k+1}-t_k} + \int_{t_k}^{t_{k+1}} \left(y(t)\right)^2  \, dt \notag\\
	& = \sum_{k \in \mathbb{Z}} \int_{t_k}^{t_{k+1}} \left(y(t)\right)^2  \, dt \notag\\
	& = \int_{-\infty}^{\infty} \left(y(t)\right)^2  \, dt \notag = \norm{y(t)}^2.\notag
	\end{align}
	Therefore, $\norm{(\mathcal{I}-\mathcal{B}_1)y(t)} = \norm{y(t)}$. Thus, $\norm{\mathcal{I}-\mathcal{B}_1} = 1$.
\end{proof}

\vspace{15pt}

\begin{lemma}\label{lemma: combined spike times are relatively separated}
Assume we have an $M$-channel TEM with parameters $\kappa$, $\delta$ and $b$, with shifts $\alpha_i\neq 0, i =1\cdots M$, and input $x(t)$ such that $|x(t)|\leq c<b$. Let $\left\lbrace\tilde{t}_k, k \in \mathbb{Z}\right\rbrace$ be the spike times generated by this $M$-channel TEM. In other words, $\left\{\tilde{t}_k, k \in \mathbb{Z}\right\}$ is the combined and ordered set of spike times generated by all channels of the TEM A$_1$, A$_2$, $\cdots$, A$_M$. Then, the spike times $\left\{\tilde{t}_k, k \in \mathbb{Z}\right\}$ are relatively separated (see Definition~\ref{def: relatively separated} in Appendix~\ref{sec: Appendix Former Results}).
\end{lemma}
\begin{proof}
Assume, without loss of generality that the channels A$_1$, A$_2$, $\cdots$ A$_M$ are ordered by spike time:
\begin{align}
    t_k^{(i)}&<t_k^{(i+1)} \quad \forall i=1\cdots M-1\notag,\\
    t_k^{(M)}&<t_{k+1}^{(1)}\notag.
\end{align}
If we denote, as in Definition~\ref{def: MTEM}, $\alpha_i$, $i=1\cdots M$ to be the shifts between two consecutively spiking machines, then a pair of consecutive spike times $\tilde{t}_k$ and $\tilde{t}_{k+1}$ will satisfy
\begin{equation*}\label{eq: spike time integral with alpha}
    \int_{\tilde{t}_k}^{\tilde{t}_{k+1}} x(u)\, du = 2\kappa\alpha_i - b\left(t_{k+1}-t_k\right), \notag
\end{equation*}
for some $\alpha_i$ that depends on the provenance of $\tilde{t}_k$ and $\tilde{t}_{k+1}$, which is determined by $k$ since different machines always spike in order (see Definition~\ref{def: MTEM}).

Now recall that $|x(t)|\leq c$, which, when substituted into~\eqref{eq: spike time integral with alpha}, yields
	\begin{align}
	    c\left(\tilde{t}_{k+1}-\tilde{t}_k\right)&\geq 2\kappa\alpha_i - b\left(\tilde{t}_{k+1}-\tilde{t}_k\right),\nonumber\\
	    \tilde{t}_{k+1}-\tilde{t}_k &\geq \frac{2\kappa\alpha_i}{b+c},\notag\label{eq:spike_seperation_lower_bound}
	\end{align}
	for some $i\in \{1,\cdots,M\}$ which depends on $k$.
	Then,
	\begin{equation*}
	    \tilde{t}_{k+1}-\tilde{t}_k \geq \frac{2\kappa\min_i(\alpha_i)}{b+c}.\notag
	\end{equation*}
	Now denote $\beta = {2\kappa\min_i(\alpha_i)}/{(b+c)}$. Note that $\beta$ is nonzero because all $\alpha_i$'s are assumed to be nonzero. Therefore, our sampling set $\left\{\tilde{t}_k, k \in \mathbb{Z}\right\}$ is relatively separated.
\end{proof}

\section*{Acknowledgements}
The authors would like to thank Ivan Dokmani\'c, Michalina Pacholska, Bernd Illing and Lara Gimena Segrelles Mun\'arriz for the fruitful discussions.

\end{appendices}

\bibliographystyle{IEEEtran}
\bibliography{main}

\end{document}